\documentclass[12pt]{article}

\usepackage{a4wide}
\usepackage[dvips]{graphicx}

\newcommand{\nn}{\nonumber}
\newcommand{\be}{\begin{equation}}
\newcommand{\ee}{\end{equation}}
\newcommand{\ba}{\begin{eqnarray}}
\newcommand{\ea}{\end{eqnarray}}
\newcommand{\ci}[1]{\cite{#1}}
\def\vk{{\bf k}_{\perp}}

\def\vd{{\bf \Delta}_\perp}
\def\vbs{{\bf b}}
\def\vb0{{\bf b}_0}
\newcommand{\LQCD}{\Lambda_{\rm{QCD}}}
\def\als{\alpha_s}
\def\ale{\alpha_{\rm elm}}
\def\mev{\,{\rm MeV}}
\def\gev{\,{\rm GeV}}

\def\xbj{x_{\rm Bj}}
\newcommand{\sla}{\hspace*{-0.20cm}/}

\newcommand{\da}{{distribution amplitude}}
\newcommand{\wf}{wavefunction}
\newcommand{\lsim}{\raisebox{-4pt}{$\,\stackrel{\textstyle
                                                         <}{\sim}\,$}}

\newcommand{\tr}[1]{{\bf #1}_\perp}

\newcommand{\req}[1]{(\ref{#1})}
\def\xb{\bar{x}}
\def\sh{\hat{s}}
\def\uh{\hat{u}}
\def\={\,=\,}
\def\eps{\epsilon}
\def\veps{\varepsilon}

\begin{document}
\thispagestyle{empty}
\begin{flushright}
WU B 05-01 \\
hep-ph/0501242\\
May 2005\\[20mm]
\end{flushright}

\begin{center}
{\Large\bf Vector Meson Electroproduction at small Bjorken-$x$ \\[0.1em]
and\\[0.4em]
Generalized Parton Distributions} \\
\vskip 15mm

S.V.\ Goloskokov
\footnote{Email:  goloskkv@thsun1.jinr.ru}
\\[1em]
{\small {\it Bogoliubov Laboratory of Theoretical Physics, Joint Institute
for Nuclear Research,\\ Dubna 141980, Moscow region, Russia}}\\
\vskip 5mm

P.\ Kroll \footnote{Email:  kroll@physik.uni-wuppertal.de}
\\[1em]
{\small {\it Fachbereich Physik, Universit\"at Wuppertal, D-42097 Wuppertal,
Germany}}\\

\end{center}

\vskip 15mm
\begin{abstract}
We analyze electroproduction of light vector mesons ($V=\rho, \phi$) at
small Bjorken-$x$ in an approach that includes the gluonic generalized
parton distributions and a partonic subprocess, $\gamma g \to (q\bar{q}) g$,
$q\bar{q}\to V$. The subprocess is calculated to lowest order of
perturbative QCD taking into account the transverse momenta of the
quark and antiquark as well as Sudakov suppressions. Our approach
allows to investigate the transition amplitudes for all kind of
polarized virtual photons and polarized vector mesons. Modelling
the generalized parton distributions through double distributions and
using simple Gaussian wavefunctions for the vector mesons, we compute 
the longitudinal and transverse cross sections at large photon
virtualities as well as the spin density matrix elements for the vector
mesons. Our results are in fair agreement with the findings of recent
experiments performed at HERA.
\end{abstract}

\centerline{(Revised version)}
\newpage
%%%%%%%%%%%%%%%%%%%%%%%%%%%%%%%%%%%%%%%%%%%%%%%%%%%%%%%%%%%%%%%%%%%%
\section{Introduction}
%%%%%%%%%%%%%%%%%%%%%%%%%%%%%%%%%%%%%%%%%%%%%%%%%%%%%%%%%%%%%%%%%%%%
Vector meson electroproduction at large photon virtuality, $Q^2$, has
always attracted a lot of theoretical interest. Its diffractive nature
as well as the interesting correlation between the $Q^2$ and the energy
dependence are challenging issues. At first traditional concepts like
vector meson dominance (see e.g.\ \ci{schuler}) or the Regge model
with its prominent Pomeron exchange (see e.g. \ci{regge})
have been exploited to analyze the electroproduction data. In 1987
Donnachie and Landshoff \ci{donn87} viewed the Pomeron as the exchange
of two gluons between the proton and the quark-antiquark pair created
by the virtual photon and which subsequently form the outgoing meson. 
Brodsky {\it et al.} \ci{bro94} treated the two-gluon exchange
contribution to electroproduction at large $Q^2$ and small
Bjorken-$x$, $\xbj$, in the framework of QCD factorization. They
showed that in their approach, known as the $\ln(1/\xbj)$
approximation, the emission and reabsorption of the gluons by the
proton can be related to the usual gluon distribution. Many variants
of the leading $\ln(1/\xbj)$ approximation can be found in the literature
which differ mainly by the treatment of the subprocess $\gamma^* g \to
Vg$, see Ref.\ \ci{kopeliovich,fra95,iva98,mrt} to name a few. These
approaches describe many features of vector meson electroproduction
quite well.

In 1996 vector meson electroproduction has been taken up by theory again.
Exploiting the new concept of generalized parton
distributions (GPD) \ci{rad96,mueller} it has been shown \ci{rad96,col96}
that, at large $Q^2$, the process
factorizes into a hard parton-level subprocess - meson 
electroproduction off partons - and soft proton matrix elements which 
represent generalized parton distributions. The process is depicted 
in Fig.\ \ref{fig:1} where also the momenta of the involved particles
are specified. It has also been shown in Refs.\ \ci{ rad96,col96} that
the process is dominated by transitions from longitudinally polarized 
photons to longitudinally polarized vector mesons ($L\to L$) at large 
$Q^2$; the amplitudes for other transitions are suppressed by inverse 
powers of $Q$. The production of vector mesons at small $\xbj$ 
( $\lsim 10^{-2}$) is controlled by gluonic GPDs where quasi on-shell 
gluons are emitted and reabsorbed by the proton. These GPDs which
represent the soft physics embodied in the proton matrix elements, are
unknown as yet and have to be modelled. 

Detailed experimental information on electroproduction of light
vector mesons in the region of small $\xbj$ is available from
HERA. Cross sections and spin density matrix elements have been
measured by H1 \ci{h1}  and ZEUS \ci{zeus98,zeus99}. Despite the sound
theoretical basis of the handbag approach not much has been done as
yet in analyzing these data within this framework. There is only the
explorative study of the longitudinal cross section for $\rho$
production performed by Mankiewicz {\it et al.} \ci{man98}. The
normalization of the cross section was however not understood in this
work. Martin {\it et al.} \ci{mrt}, on the other hand, started from
the $\ln{(1/\xbj)}$ approximation and estimated effects due to the
replacement of the gluon distribution by the corresponding GPD.
Here, in this work we attempt a complete and systematic analysis of
the available electroproduction data at small $\xbj$. In order to
analyze the spin density matrix elements of the vector mesons 
we also calculate the amplitudes for transitions from transversely 
polarized photons to transversely and longitudinally polarized vector 
mesons ($T\to T$ and $T\to L$). We allow for quark 
transverse momentum and take into account Sudakov suppressions. As it
will turn out this approach leads to the correct normalization of the
cross sections at finite but large $Q^2$. Infrared singularities which
occur for the $T\to T$ transition amplitude in collinear approximation 
\ci{man99}, are also regularized in our approach although in an
admittedly model-dependent way. 

The plan of the paper is the following: A kinematical prelude and
the handbag amplitude are presented in Sect.\ 2. The amplitudes
for the subprocess $\gamma^* g\to V g$ are discussed in Sect.\ 3 to
leading order of perturbative QCD and including transverse momenta of
the quarks and antiquarks making up the meson. The impact parameter 
representations of the full handbag amplitudes for electroproduction  of 
vector mesons are presented in Sect.\ 4. The following section, 
Sect.\ 5, is devoted to the construction of the GPDs. Numerical
results, obtained from the handbag approach, for the cross sections 
of vector-meson electroproduction and for the vector meson's spin 
density matrix elements are compared to recent experimental results in 
the small $\xbj$ region in Sects.\ 6 and 7, respectively. In the next 
section, Sect.\ 8, we discuss the helicity correlation $A_{LL}$ and
the role of the GPD $\widetilde{H}$ and summarize in Sect.\ 9. 

%%%%%%%%%%%%%%%%%%%%%%%%%%%%%%%%%%%%%%%%%%%%%%%%%%%%%%%%%%%%%%%%%%%%%%%    
\section{The handbag factorization}
%%%%%%%%%%%%%%%%%%%%%%%%%%%%%%%%%%%%%%%%%%%%%%%%%%%%%%%%%%%%%%%%%%%%%%%
\begin{figure}[t]
\begin{center}
\includegraphics[width=.38\textwidth,bb=112 536 350 715,clip=true]{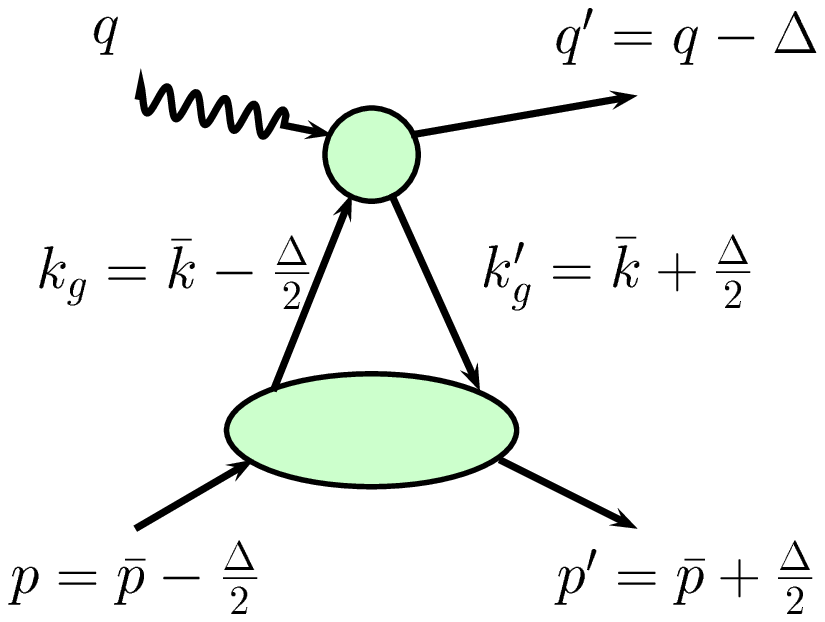}
\end{center}
\caption{The handbag-type diagram for meson electroproduction off
protons. The large blob represents a GPD while the small one
stands for meson electroproduction off partons. The momenta of the
involved particles are specified.}
\label{fig:1}
\end{figure}
We will work in a photon-proton center of mass (c.m.) frame, see Fig.\
\ref{fig:1}, in a kinematical situation where 
\be
W^2 \= (p+q)^2\,,
\ee
and the virtuality of the
incoming photon, $q^2=-Q^2$, are large while Bjorken's variable, 
\be
\xbj\=Q^2/(2p\cdot q)\,,
\ee
is small ($\xbj\lsim 10^{-2}$). We also assume the square of
the momentum transfer, $\Delta=p'-p$, to be much smaller
than $Q^2$.
 
In light-cone components, defined by  $a^\pm=(a^0 \pm a^3)/\sqrt{2}$
and $a \cdot b=a^{+} b^{-}+a^{-} b^{+} - {\bf a}_\perp {\bf b}_\perp$,
the momenta of the protons and the photon read 
\ba
\label{ml}
p&=&\left[(1+\xi)\bar{p}^{+},\frac{m^2+\vd^2/4}{2(1+\xi)\bar{p}^+}\,,
                                       -\frac{\vd}{2}\right]\,, \nn\\
p'&=&\left[(1-\xi) \bar{p}^+,
  \frac{m^2+\vd^2/4}{2(1-\xi)\bar{p}^+}\,,\phantom{-} 
                                          \frac{\vd}{2}\right]\,,\nn\\ 
q&=&\left[\eta (1+\xi)\bar{p}^+,\frac{-Q^2+\vd^2/4}{2\eta
    (1+\xi)\bar{p}^{+}}\,, \frac{\vd}{2}\right]\,,
\ea
where $\eta$ equals $-\xbj$ up to corrections of order $m^2/Q^2$ and
$\vd^2/Q^2$. Here, $m$ denotes the mass of the proton.
The average proton momentum is defined by
\be
\bar{p}\= \frac12\, (p+p')\,,
\label{eq:avgp}
\ee
and the skewness parameter $\xi$ by
\be 
\xi \= \frac{(p-p')^+}{(p + p')^+}\,.
\ee
In the photon-proton c.m.\ frame and for small $\xbj$, the skewness parameter is
related to Bjorken-$x$ by 
\be 
\xi \= \frac{\xbj}{2-\xbj} \simeq \xbj/2\,.
\label{skew}
\ee
For Mandelstam $t$, given by   
\be
t \=\Delta^2 \=-\frac{4\xi^2 m^2 + \vd^2}{1-\xi^2}\,,
\ee
a minimal value is implied by the positivity of $\Delta_\perp^2$  
\be 
     -t_{\rm min} \= 4m^2 \frac{\xi^2}{1-\xi^2}\,.
\label{tmin}
\ee
Since we are interested in the region of small Bjorken-$x$ and, hence,
small skewness we will use $t_{\rm min}\simeq 0$ in the following.
We also will neglect the proton and meson ($m_V$) masses in the
kinematics.

Let us now consider the dynamics of vector meson electroproduction in
the kinematical regime specified above. The dominant contribution in
this kinematical region comes from the emission and reabsorption of
collinear gluons from the protons accompanied by $\gamma^* g \to Vg$
scattering \ci{rad96}. The neglect of an analogous quark contribution
is justified by the fact that, at small $\xbj$, partons with small 
momentum fractions dominantly participate in hard meson electroproduction. 
Since, at small $-t$, the GPDs are expected to reflect the magnitudes of 
the usual parton distributions the gluon contribution should outweigh
the quark one. This is in particular the case for electroproduction of
$\phi$ mesons where only the small strange quark content of the proton
is probed. Even for the production of $\rho$ mesons the gluonic
contribution seems to be still sizeable for $\xbj$ as large as 0.1 as is
indicated by the ratio of $\phi$ and
$\rho$ electroproduction cross sections~\ci{diehl04}.    

The momenta of the gluons which, approximately, are collinearly emitted or
absorbed by the protons, are parameterized as 
\ba
k_g &=& \left[ (\xb+\xi)\bar{p}^+,\frac{\vd^2}{8(\xb+\xi)\bar{p}^+},-\vd/2 \right]\,, \nn\\ 
k_g'&=&  \left[(\xb-\xi) \bar{p}^+,\frac{\vd^2}{8(\xb-\xi)\bar{p}^+}\,,\;\;\vd/2\right]\,.
\label{parton-mom}
\ea
In general the partons may have small virtualities of the order of $\vd^2$.
As usual we have introduced an average parton momentum 
\be
\bar{k} \= \frac12 (k_g+k_g')\,,
\ee
and an average momentum fraction 
\be
\xb\= \bar{k}^+/\bar{p}^+\,.
\ee 

In order to facilitate comparison with other work we also provide the
relations between the variables $\xb$ and $\xi$ and the usual
Mandelstam variables for the hard subprocess. They read
($\hat{t}\simeq 0$) 
\begin{eqnarray}
\sh&=&(q+k_g)^2 \simeq \frac{\xb-\xi}{2\xi}\, Q^2\,, \nn\\ 
\uh&=&(q-k_g')^2  \simeq  -\frac{\xb+\xi}{2\xi}\, Q^2\,,   
\label{man}
\end{eqnarray}
and are valid at large $Q^2$ and small $\xi$. 

Radyushkin has calculated the asymptotically leading handbag
contribution to meson electroproduction at small $\xbj$ \ci{rad96}. 
As he showed this contribution involves $L\to L$ transitions. 
Leaving aside for the time being a potential breakdown of
factorization, Radyushkin's result can straightforwardly be
generalized to other transitions \ci{man99,hanwen}. The crucial 
point in the derivation of the handbag amplitude is the use of 
light-cone gauge for the gluon field, $n\cdot A^a=0$, where
\be
n\=\left[0,1,{\bf 0}_\perp \right]\,,
\ee
and $a$ is a color label. This gauge allows to express the gluon field 
by an integral over the gluon field strength tensor $G^a_{\nu\nu^\prime}$
\ci{rad96,kog70} (the limit $\tilde{\varepsilon}\to 0$ is to be understood) 
\begin{equation}
A^a_\nu(z) \= n^{\nu^\prime} \int_0^\infty d\sigma 
                 {\rm e}^{-\tilde{\veps}\sigma}\, G^a_{\nu\nu^\prime}(z+\sigma n)\,.
\end{equation}
With the help of this expression one can replace the products of
fields appearing in the perturbatively calculated amplitude for 
$\gamma^* p \to Vp$ by
\ba
 A^{a\rho}(0)\, A^{a'\rho^\prime}(\bar{z}) &=& \frac{\delta^{aa'}}{N_c^2-1}\,
                  \sum_{\lambda,\lambda^\prime=\pm 1}
                  \,\epsilon^\rho (k_g,\lambda)\, \epsilon^{*\rho^\prime}
                                               (k_g^\prime,\lambda^\prime)
                   \int d\sigma\,
                   d\sigma^\prime\, {\rm e}^{-\tilde{\varepsilon}\sigma-
                              \tilde{\varepsilon}^\prime\sigma^\prime} 
                                                              \nonumber\\
                 &\times& n^\omega\, n^{\omega^\prime}\,
                  G^a_{\nu\omega}(\sigma^\prime n)\,  
                       G^a_{\nu^\prime \omega^\prime}(\bar{z}+\sigma n)\,        
                   \epsilon^{*\nu}(k_g,\lambda)\,
                  \epsilon^{\nu^\prime}(k_g^\prime,\lambda^\prime)\,,
\label{glue-replace}
\end{eqnarray}
where we have also made a helicity projection for the gluons. The use
of the approximation \req{parton-mom} for the gluon momenta forces the
relative distance of the fields on the light cone 
$z\to \bar{z}=[0,z_-,{\bf 0}_\perp]$. The vectors $\eps(k_g,\lambda)$
and $\eps(k_g^\prime,\lambda^\prime)$ specify the polarization of the 
(on-shell) gluons, the corresponding momenta, $k_g$ and $k_g^\prime$, are 
defined in Eq.\ (\ref{parton-mom}). The first set of polarization
vectors in (\ref{glue-replace}) is to be used to contract the hard 
scattering kernel leading to gauge invariant parton-level helicity
amplitudes ${\cal H}^V_{\mu^\prime\lambda^\prime,\mu\lambda}$ for 
$\gamma^* g\to Vg$ ($\mu$ and $\mu^\prime$ denote the helicities of
$\gamma^*$ and $V$, respectively). The contraction of the field
strength tensors with the second set of polarization vectors leads to
\cite{DFJK3}
\ba
\lefteqn{n^\omega n^{\omega^\prime}\, G_{\nu \omega}(\sigma^\prime n)  
           \, G_{\nu^\prime\omega^\prime}(\bar{z}+\sigma n) 
             \,\epsilon^{*\nu}(k_g,\lambda) 
                      \, \epsilon^{\nu^\prime}(k_g^\prime,\lambda^\prime)=} \nn\\
  && \hspace{10mm} \frac12\,n^\omega n^{\omega^\prime}\, 
             G_{\nu\omega}(\sigma^\prime n)\,  
                        G_{\nu^\prime \omega^\prime}(\bar{z}+\sigma n) \,
         \left[\, (-g^{\nu\nu^\prime}_\perp + 
                          \lambda\,i\epsilon_\perp^{\nu\nu^\prime})\,
                          \delta_{\lambda \lambda^\prime} 
                    -t_\perp^{\nu\nu^\prime}\, 
                       \delta_{\lambda -\lambda^\prime}  \right]\,,
\label{decomp}
\ea
where 
\ba
g^{11}_\perp &=& g^{22}_\perp = -\epsilon_\perp^{12}
=\epsilon_\perp^{21}= -t_\perp^{11}=  t_\perp^{22}=-1 \nn\\
t_\perp^{12} &=&  t_\perp ^{21} = i \lambda
\label{trans-tensors}
\ea
while all other components of these tensors are zero. That only the
transverse components in the contraction remain is a consequence of
the chosen light-cone gauge and of the fact that the polarization
vectors have zero plus components in the c.m.\ frame we are
working. 

Proton matrix elements of the gluon helicity non-flip contributions
$g_\perp^{\mu\mu'}$ and $i\epsilon_\perp^{\mu\mu'}$ in \req{decomp} 
define the unpolarized, $H^g(\xb,\xi,t)$ and $E^g(\xb,\xi,t)$, and 
the polarized, $\widetilde{H}^g(\xb,\xi,t)$ and 
$\widetilde{E}^g(\xb,\xi,t)$, gluon GPDs, respectively \ci{rad96,mueller}.
The proton matrix elements of these gluon field operators are related to
the GPDs by 
\ba
\lefteqn{
\langle p'\nu'|\sum_{a,a'} A^{a\rho}(0)\, A^{a'\rho'}(\bar{z})|p\nu\rangle  
      \= \frac12  \sum_{\lambda=\pm 1} \eps^\rho(k_g,\lambda)\, 
                 \eps^{*\rho^\prime} (k_g',\lambda')}  \\
  &\times& \int_0^1 
         \frac{d\xb}{(\xb+\xi- i{\veps})(\xb-\xi + i{\veps})}\,
           {\rm e}^{-i(\xb-\xi) p\cdot\bar{z}} \nn\\
   &\times& \left\{\frac{\bar{u}(p'\nu')\,n\sla\, u(p\nu)}{2\bar{p}\cdot
                                                 n}\;H^g(\xb,\xi,t)
            + \frac{\bar{u}(p'\nu')\,i\, \sigma^{\alpha\beta}\,n_\alpha
         \Delta_\beta\, u(p\nu)}{4m\, \bar{p}\cdot n}\;{E}^g(\xb,\xi,t)\right.\nn\\   
      &+& \left.\lambda \frac{\bar{u}(p'\nu')\,n\sla\gamma_5\, u(p\nu)}{2\bar{p}\cdot
                                                 n}\;\widetilde{H}^g(\xb,\xi,t)    
          + \lambda \frac{\bar{u}(p'\nu')\,n\cdot\Delta\,\gamma_5\,
         u(p\nu)}{4m\, \bar{p}\cdot n}\;\widetilde{E}^g(\xb,\xi,t)\right\}\,.\nn
\label{eq:matrix}
\ea
Working out the spinor products one sees that for proton helicity non-flip 
the linear combinations \cite{DFJK3}
\be
     H^g(\xb,\xi,t) -\frac{\xi^2}{1-\xi^2}\, E^g(\xb,\xi,t)    
\label{HE}
\ee
and 
\be 
\widetilde{H}^g(\xb,\xi,t) -\frac{\xi^2}{1-\xi^2}\, \widetilde{E}^g(\xb,\xi,t)
\label{HE-tilde}
\ee
occur. Since we are interested in small $\xi$ the $E^g$ and 
$\widetilde{E}^g$ terms can safely be neglected in the  expressions 
\req{HE} and \req{HE-tilde}. For proton helicity flip, on the other 
hand, $H^g$ and $\widetilde{H}^g$ do not contribute but only
\be 
    -\kappa\, \frac{\sqrt{-t}}{2m}\,\frac{1}{{1-\xi^2}} \;
                  E^g \;(\,2\nu\xi\widetilde{E}^g\,)\,,
\ee
where $\kappa$ is a phase factor reading
\be
 \kappa = \frac{\Delta^1 + i\Delta^2}{|{\bf \Delta_\perp}|}\,,
\ee
for proton momenta of the form (\ref{ml}).

The gluon helicity flip contribution in (\ref{decomp}) which defines
four more GPDs \ci{diehl}, will be neglected in the following since it
is strongly suppressed at small $-t$.  The mismatch between the
proton and gluon helicities in the proton matrix elements
has to be compensated by orbital angular momentum. For each unit of it a
factor $\sqrt{-t}/m$ is picked up \ci{diehl,diehl03}. Further
suppression comes from the subprocess amplitudes which behave as 
\be
{\cal H}^V_{\mu'\lambda',\mu\lambda} \sim 
                  \Big(\sqrt{-t}/Q\Big)^{|\mu-\lambda-\mu'+\lambda'|}\,, 
\label{eq:smallt}
\ee
at small $-t$ and from the fact that the amplitude 
${\cal H}^V_{0-\lambda,\mu\lambda}$ vanishes for $\mu=\pm 1$ \ci{hanwen}.  

Combining all this, we finally obtain the helicity amplitudes for
electroproduction of vector mesons~\footnote{We note in passing that
  our helicities are light-cone helicities which naturally occur in
  the handbag approach. The difference to the usual c.m.\ frame
  helicities is of order $m\sqrt{-t}/W^2$ \ci{diehl} and can be
  ignored in the kinematical region of interest in this work.}:   
\ba
{\cal M}_{\mu'+,\mu +} &=& \frac{e}{2}\, {\cal C}_V\,
         \int_0^1 \frac{d\xb}
        {(\xb+\xi)
                  (\xb-\xi + i{\veps})}\nn\\
        &\times& \left\{\, \left[\,{\cal H}^V_{\mu'+,\mu +}\,
         + {\cal H}^V_{\mu'-,\mu -}\,\right]\, 
                                   H^g(\xb,\xi,t) \right. \nn\\
       &+& \hspace{2.5mm}\left. \left[\,{\cal H}^V_{\mu'+,\mu +}\,
            -   {\cal H}^V_{\mu'-,\mu -}\,\right]\,  
                        \widetilde{H}^g(\xb,\xi,t)\, \right\}\,,
\label{amp-nf}
\ea
for proton helicity non-flip (explicit helicities are labelled by
their signs) and for helicity flip
\ba 
{\cal M}_{\mu'-,\mu +} &=& -\frac{e}{2} \,{\cal C}_V \kappa\,
\frac{\sqrt{-t}}{2m} \; 
   \int_0^1 \, \frac{d \xb}{(\xb+\xi) 
             (\xb-\xi+i{\veps})} \nn\\
  &\times& \left\{\,\left[\,{\cal H}^V_{\mu'+,\mu +} + \,{\cal H}^V_{\mu'-,\mu -}\right]
                    E^g(\xb,\xi,t) \right. \nn\\     
        &+& \hspace{2.5mm} \left.\left[\,{\cal H}^V_{\mu'+,\mu +} - 
                    \,{\cal H}^V_{\mu'-,\mu -}\right]
                   \xi\widetilde{E}^g(\xb,\xi,t) \right\}\,. 
\label{amp-f}
\ea
The subprocess amplitudes, ${\cal H}^V$, are functions of $Q^2$,
$\xb$, $\xi$ and $t$. The flavor weight factors, ${\cal C}_V$, read
for $\rho$ and $\phi$ mesons
\be
{\cal C}_\rho \= \frac1{\sqrt{2}}\, (e_u  - e_d ) \= 1/\sqrt{2}\,;
\qquad {\cal C}_\phi \= e_s  \= -1/3\,.
\ee 
where $e_i$ denotes the quark charge in units of the positron charge
$e$. The remaining helicity amplitudes are obtained with the help of
parity invariance
\be
  {\cal M}_{-\mu' -\nu',-\mu -\nu} \= (-1)^{\mu-\nu-\mu'+\nu'}\; 
                                       {\cal M}_{\mu' \nu',\mu \nu}\,. 
\label{parity}
\ee
An analogous relation holds for the subprocess amplitudes.

%%%%%%%%%%%%%%%%%%%%%%%%%%%%%%%%%%%%%%%%%%%%%%%%%%%%%%%%%%%%%%%
\section{The partonic subprocess $\gamma^* g \to Vg$}
\label{sec:sub}
%%%%%%%%%%%%%%%%%%%%%%%%%%%%%%%%%%%%%%%%%%%%%%%%%%%%%%%%%%%%%%%%
\begin{figure}[bt]
\begin{center}
\includegraphics[width=0.48\textwidth,
  bbllx=85pt,bblly=365pt,bburx=518pt,bbury=653pt]{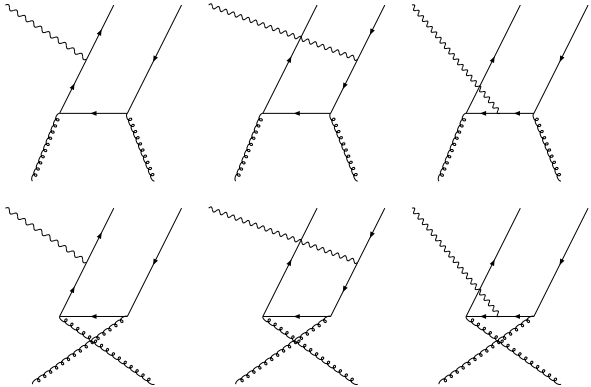}
\end{center}
\caption{Lowest order Feynman graphs for the subprocess $\gamma^* g\to
Vg$. }
\label{fig:2}
\end{figure}

The parton-level amplitudes for the subprocess $\gamma^* g\to Vg$ are 
calculated from the Feynman graphs shown in Fig.\ \ref{fig:2}; 
the outgoing $q\bar{q}$ pair is to be combined into the vector meson
regarding its quantum numbers. This is conveniently done by means of
a covariant spin \wf. As is well-known from analyses of hadron form 
factors at large momentum transfer, leading-twist
perturbative calculations are instable in the end-point regions since
the contributions from large transverse separations, $\vbs$, of quark
and antiquark forming the meson are not sufficiently suppressed. In 
order to eliminate that defect Li and Sterman \cite{li92} retained the
quark transverse degrees of freedoms and took into account Sudakov
suppressions. Including, in addition, meson \wf s with their 
intrinsic transverse momentum dependence instead of \da{}s \cite{jak93}, 
the perturbative contributions to form factors can reliably and 
self-consistently be calculated, the end-point regions are strongly
damped.
 
Since the subprocess $\gamma^*g\to Vg$ bears resemblance to the
meson form factors it is tempting to apply this so-called modified
perturbative approach also here in order to suppress the contributions
from the soft end-point regions  and, simultaneously, to regularize
this way infrared divergencies that may occur in the $T\to L$  and
$T\to T$ amplitudes \ci{man99}. The modified perturbative approach 
applied to the subprocess, is, to some extent, similar to the
mechanism proposed in Ref.\ \cite{fra95} for the suppression of the 
leading-twist gluon contribution to hard meson electroproduction. It 
is however to be stressed that in Ref.\ \ci{fra95} the leading $\ln(1/\xbj)$
approximation of Ref.\ \ci{bro94} has been utilized.

Let us now turn to the description of the soft $q\bar{q}\to V$ transtion
matrix element. We start from a frame where the hadron rapidly moves
along the 3-direction ($q'=[0,q'{}^-,\tr{0}]$ with $q'{}^-\simeq
Q/(2\sqrt{\xi})$). This frame is termed the hadron-out one.  
The momenta of the quark and the antiquark which form the valence Fock
state of the meson, are parameterized as 
\be
q_1^\mu\= \tau q'{}^\mu + k_1^\mu\,, \qquad  
                                  q_2^\mu\= \bar{\tau} q'{}^\mu + k_2^\mu\,.
\label{quark-mom}
\ee
where 
\be
k_1\=[k_1^+, 0, {\bf k}_{1\perp}]\,, \qquad 
                             k_2\=[k_2^+, 0, {\bf k}_{2\perp}]\,.
\ee     
The variables $\tau$ and $\bar{\tau}$ are the usual fractions of the
light-cone minus-component of the meson's momentum the constituents 
carry. Momentum conservation provides the constraints
\be
\bar{\tau}\=1-\tau\,, \qquad {\bf k}_{2\perp}\=- {\bf k}_{1\perp}\equiv
- \vk\,.
\ee
It can be shown \ci{leutwyler} that the variables $\tau, \bar{\tau}$
and ${\bf k}_{\perp}$ are invariant under all kinematical Poincare
transformations, i.e.\ under boosts along and rotations around the
3-direction as well as under transverse boosts. Moreover - and this is
an important point - the light-cone \wf{} associated with the valence Fock
state, $\Psi_V=\Psi_V(\tau,\vk)$, is independent of the hadron's
momentum and is invariant under these kinematical transformations too.
The light-cone \wf{} may differ for longitudinally and transversally
polarized vector mesons~\ci{CZ}.

As is customary in the parton approach we neglect the binding
energy. That possibly crude approximation can be achieved by putting
the individual $k_j^+$-components to zero. In fact starting from a
parameterisation of the various momenta in the meson's rest frame and
boosting to the hadron-out frame, one sees that the $k_j^+$-components
are of order $m_V^2/q^{\prime -}$. The plus-component of the 
difference of the momenta 
\be
K\=\frac12\,(k_1-k_2)\,,
\label{rel-mom}
\ee
is zero with this choice of $k_j^+$ components and, hence, 
\be
K\=[0,0,\vk]\,,
\label{K-def}
\ee
and $q^\prime \cdot K=0$. The quarks are treated as massless in the
hadron-out frame; they are not strictly on-shell.
   
It is convenient to couple the spinors representing quark and
antiquark in a covariant spin wave function for the vector meson. 
The Dirac indices of it (omitted for convenience)
are to be contracted with the corresponding ones of the hard
scattering kernel (see below). For the construction of the spin wave
function we adapt the method presented in Ref.\ \ci{koerner} (see also
\ci{guberina}) straightforwardly to vector mesons. The product of
spinors $v(q_2) \bar{u}(q_1)$ is boosted to the hadron's rest frame,
coupled there into the quantum numbers of the vector meson and boosted
back to hadron-out frame. Separating terms with and without $K$ and
neglecting terms $\propto K^2$, one arrives at
\be
\Gamma_V \=  \Gamma_{V0} + \Delta \Gamma_{V\alpha} K^\alpha\,,
\label{spin}
\ee
where
\be
 \Gamma_{V0} \= \frac1{\sqrt{2}}\, \left( q\sla{}' + m_V \right)
 \eps\sla_V\,,
\qquad
\Delta \Gamma_\alpha \= \frac1{M_V}\, \left\{\Gamma_{V0},\gamma_\alpha \right\}\,.
\label{spin01}
\ee
The polarization state of the meson is described by the vector $\eps$. 
The soft physics parameter $M_V$ is of order $m_V$; its model
dependence results from the specific treatment of the quarks in the
meson's rest frame. In the following we will use $M_V=m_V$ for simplicity
but we will comment on other choices of it.

Since the anticommutator $\left\{\Gamma_{V0},q\sla'\right\}$ is zero the
4-vector $K$ is only determined up to a multiple of the meson
momentum. This property can be used to identify $K^\mu$, given in Eq.\
\req{K-def}, with the quark-antiquark relative momentum
\be
        K^\mu \to \frac12 (q_1-q_2)\,,
\label{K-rel-mom}
\ee
where the parton momenta, $q_i$, are defined in Eq.\ \req{quark-mom}.
This choice, although not forced, is very convenient. Its
main advantage is that, $K^\mu$ now represents one unit of orbital
angular momentum in a covariant manner~\ci{koerner}. As discussed in
this article, the relative momentum \req{K-rel-mom} is a 4-transverse
vector which are defined by $K^\mu_\perp = K^\mu - q^{\prime}\cdot
K/m_V^2\, q^{\prime\mu}$. In the hadron-out frame and
up to corrections of order $m_V^2/q^{\prime -}$, $K_\perp = K$. In the
meson's rest frame on the other hand, clearly $K_\perp = (0,\vec{k})$, 
and one has an object transforming as a 3-vector under the three-dimensional
rotation group $O(3)$. 

One of the basis ingredients of the
hard scattering picture is the collinear approximation which says that
all constituents move along the same direction as their parent hadron
up to a scale of the order of the Fermi motion $\langle \vk^2\rangle$
which typically amounts to a few 100 MeV. The (nearly) collinear
kinematics justifies an expansion of the spin \wf{} upon a power
series in $\vk$ or, in order to retain a covariant formulation, in
$K^\mu$. Up to terms linear in $K^\mu$ this expansion is given above
for vector mesons. 

The transformation from the hadron-out frame to our c.m.\ frame where the 
meson momentum has a transverse component $-\vd/2$, is executed by a
transverse boost (cf.\ e.g.~\cite{bro89}) that leaves the minus
component of any momentum vector $a$ unchanged, and which involves a
parameter $d^-$  and a transverse vector ${\bf d}_\perp$, is defined as
\be
[\,a^+, a^- , {\bf a}_\perp] 
\qquad\longrightarrow\qquad
[\,a^{+} -\frac{{\bf a}_\perp\cdot{\bf d}_\perp}{d^-}
+\frac{a^-\,{\bf d}_\perp^{\,2}}{2\,(d^-)^2}\,,\,a^{-}\,,
      {\bf a}_\perp-\frac{a^-}{d^-}\,{\bf d}_\perp] \,.
\ee 
The transverse boost is one of the kinematical Poincare transforms
that leaves the hadron \wf{} invariant. Taking for the parameters
$d^-= q'{}^-$ and ${\bf d}_\perp=\vd/2$, we readily find from
\req{quark-mom} the expressions for the quark momenta and the relative
momentum in the c.m.\ frame. 
 
Provided quark transverse momenta are taken into account, the 
general structure of the $\gamma^* g \to Vg$ amplitude is 
\be
{\cal H}^V \= \int \frac{d\tau d^2\vk}{16\pi^3}\,
\Psi_V (\tau,k_\perp^2)\, {\rm Tr}\left[\Gamma_V\, T_H\right]\,.
\label{structure}
\ee 
The hard scattering kernel, $T_H=T_H(\tau,\xb,Q^2,K,t)$, can
be written as follows  
\be
T_H\= T_0(\tau,\xb,Q^2,k_\perp^2,t) + 
                   \Delta T_{\mu}(\tau,\xb,Q^2,k_\perp^2,t) K^\mu\, +\dots.
\label{exp-T}
\ee
Terms $\propto K^\mu K^\nu$ (and higher) in  the nominator are neglected
while, in the spirit of the modified perturbative approach~\ci{li92}, 
the $k_\perp^2$ terms in the denominators are kept.   
Inserting Eq.\ \req{exp-T} as well as the spin \wf{} \req{spin} 
into Eq.\ \req{structure}, one obtains  
\ba
{\cal H}^V &=& \int \frac{d\tau
  d^2\vk}{16\pi^3}\,\Psi_V(\tau,k^2_\perp)\, 
            {\rm Tr}\left[ \Gamma_{V0}\, T_0 + \Delta \Gamma_{V\alpha}\, 
                                            T_0\, K^\alpha \right.\nn\\
    && \left. \hspace{1cm} + \Gamma_0 \,\Delta T_\beta\, K^\beta + 
         \Delta \Gamma_{V\alpha}\, \Delta\, T_\beta\, 
                                           K^\alpha\, K^\beta + \dots\right]\,.  
\ea
Obviously, the terms $\propto K^\mu$ integrate to zero while the 
$K^\mu K^\nu$ term survives the $k_\perp$-integration . Hence,
\be
{\cal H}^V\=\int d\tau \frac{d k^2_\perp}{16\pi^2} \Psi_V(\tau,k^2_\perp) 
    {\rm Tr} \left\{\Gamma_{V0} T_0 
       - \frac12 k^2_\perp g_\perp^{\alpha\beta}\, \Delta\Gamma_{V\alpha}\, 
            \Delta T_\beta + \dots\right\}\,,
\label{amp-mpa}
\ee
where $g_\perp$ is the transverse metric tensor defined in
Eq.\ \req{trans-tensors}. In order to simplify matters we only take into
account the first non-zero term in this expansion for each amplitude,
i.e.\ we neglect any correction of order $m_V$ or $k^2_\perp$ to its
leading term~\footnote{
   Note that the hard scattering kernel $T_H$ does not depend on the
   vector meson mass; it occurs through the spin wave function.}.
As we said above we however retain the $k^2_\perp$ terms in the denominators
of the propagators. Morever, any $t$ dependence of the subprocess 
amplitudes is ignored except the factors of $\sqrt{-t}$ required by
angular momentum conservation. This is justified in the small $t$
region we are interested in.

For longitudinally polarized vector mesons the first term in
Eq.\ \req{amp-mpa} contributes, the other term  represents a
$k^2_\perp$ correction to it which we, according to our strategy,
neglect as well as all other terms indicated by the ellipses. For
transversely polarized mesons, on the other hand, the first term in
Eq.\ \req{amp-mpa} disappears since the number of $\gamma$ matrices in
the trace is odd~\footnote{
 We remind the reader that for longitudinally polarized vector mesons 
 $\eps(0)=q'/m_V$ up to corrections of order $m_V/q^{\prime -}$.}. 
The second term in Eq.\ \req{amp-mpa}, $\propto k^2_\perp$,
contributes in this case; it scales as $\propto k^2_\perp/(m_V Q)$,
see Eq.\ \req{spin01}. Combining this property with the behavior of the 
subprocess amplitudes near the forward direction \req{eq:smallt} and
utilizing the fact that $\langle k^2_\perp\rangle^{1/2}/m_V$ is of
order 1, the various photon-meson transitions respect the following hierarchy 

\ba
{\rm L\to L}:&& \qquad {\cal H}^V_{0\,\lambda,0\,\lambda} \propto 1\,, \nn\\
{\rm T\to L}:&& \qquad {\cal H}^V_{0\,\lambda,+\lambda} \propto
                                        \frac{\sqrt{-t}}{Q}\,, \nn\\
{\rm T\to T}:&& \qquad {\cal H}^V_{+\lambda,+\lambda} \propto 
                          \frac{\langle k^2_\perp\rangle^{1/2}}{Q}\,, \nn\\
{\rm L\to T}:&& \qquad {\cal H}^V_{+\lambda,0\,\lambda} \propto
     \frac{\sqrt{-t}}{Q}\, \frac{\langle k^2_\perp\rangle^{1/2}}{Q}\,, \nn\\
{\rm T\to -T}:&& \qquad {\cal H}^V_{-\lambda,+\lambda} \propto 
                    \frac{-t}{Q^2}\, \frac{\langle k^2_\perp\rangle^{1/2}}{Q}\,. 
\label{hierarchy}
\ea
This hierarchy propagates to the proton non-flip amplitudes for the full
process and justifies the neglect of $L\to T$ and $T\to -T$
transitions in the analysis. The amplitudes for proton helicity flip
have an extra factor $\sqrt{-t}/m$, see \req{amp-f}. Our interest in
this work is focussed on unpolarized protons. In the corresponding
cross sections there is no interference between flip and non-flip
amplitudes. Hence, proton flip is suppressed by a factor of $t$ and 
since there is no theoretical or phenomenological indication that 
$|E^g|$ is much larger than $H^g$ \ci{diehl03}, neglected by us. 
Information on the proton flip amplitudes may be extracted from data
on meson electroproduction with polarized protons. As a last 
simplification we neglect contributions from $\widetilde{H}^g$ in the
evaluation of the amplitudes. Since in the forward limit $\xi, t\to 0$, 
$H^g$ and $\widetilde{H}^g$ reduce to $\xb g(\xb)$ and  $\xb\Delta g(\xb)$, 
respectively, it is plausible to expect that the relative size of 
$\Delta g$ and $g$ is reflected in that of $\widetilde{H}^g$ and 
$H^g$ at small $\xi$ and $-t$. Since $|\Delta g(\xb)|$ is much smaller 
than $g(\xb)$ the contribution from  $\widetilde{H}^g$ can safely be
neglected. The model GPDs we are going to construct in Sect.\ 
\ref{sec:gpd} do indeed respect this assertion. As a consequence of 
parity invariance, see \req{amp-nf} and \req{parity}, there is anyway
no contribution from the GPD $\widetilde{H}^g$ to the most important
amplitude, $L\to L$. Care is required for observables for which the 
contribution from $H^g$ partially if not totally cancels. An example 
of such an observable is the correlation of the electron and proton
helicities. We will comment on this observable in Sect.\ \ref{sec:all}.

The hard scattering amplitudes for the three helicity configurations
we keep in our analysis are to be calculated from the Feynman graphs
shown in Fig.\ \ref{fig:2}. The results for the relevant sums and
differences of positive and negative gluon helicities can be cast into
the following form
\be
{\cal H}^V_{\mu'+,\mu +}\,\pm\, {\cal H}^V_{\mu'-,\mu -}\,  
                                       \= \,\frac{8\pi \als(\mu_R)}
           {\sqrt{2N_c}} \,\int_0^1 d\tau\,\int \frac{d^{\,2} \vk}{16\pi^3}
            \Psi_{V\mu^\prime}(\tau,k^2_\perp)\; 
                  (\xb^2-\xi^2)\,f_{\mu^\prime\mu}^\pm D\,,
\label{tt-ji}
\ee
where the product of propagator denominators reads
\ba
D^{-1} &= &\left(k^2_\perp + \bar{\tau}\,Q^2\right)\;  
                        \left(k^2_\perp + \tau\,Q^2 \right)\; \nn\\\ 
   &\times&\left(k^2_\perp - \bar{\tau}\, (\xb-\xi)\, Q^2/(2\,\xi) -i{\veps}\right)\;  
    \left( k^2_\perp+ \bar{\tau}\,(\xb+\xi)\,Q^2/(2\,\xi) \right)\;\nn\\
      &\times&          \left(k^2_\perp + \tau\,(\xb+\xi)\, Q^2/(2\,\xi) \right)\;  
                \left(k^2_\perp -\tau\,(\xb-\xi)\,Q^2/(2\,\xi) -i{\veps} \right)\,.
\label{propagator}
\ea
Here, $N_C$ denotes the number of colors. The functions $f^\pm_{\mu'\mu}$ read  
\ba
f^+_{00} &=& \phantom{-}Q^{11}\,(\xb^2-\xi^2)\, 
                      \frac{\tau^2\bar{\tau}^2 }{4\,\xi^4} \,,\nn\\[0.3em]
f^+_{0+} &=&  Q^{10} \, \sqrt{\frac{-t}{2}} \; (\xb^2-\xi^2)^{1/2}\,
                         \frac{\tau^2\bar{\tau}^2}{2\xi^3}\,, \nn\\[0.3em]    
f^+_{++}&=& -\frac{k^2_\perp}{m_V}\, Q^{10}\, 
           \frac{\tau\bar{\tau}}{8\,\xi^4}\,
           [\xb^2-\xi^2- 2\tau\bar{\tau}\,(\xb^2+\xi^2)]\,,\nn\\ [0.3em]  
f^-_{++} &=&\phantom{-} \frac{k^2_\perp}{m_V}\, Q^{10}\, 
           \frac{\tau^2\bar{\tau}^2}{2\,\xi^3}\,\xb\,, \nn\\[0.3em]
f^-_{00} &=& f^-_{0+} \=  0\,.
\label{hard-amp-ji}
\ea
Following Ref.\ \ci{li92}, we retain $k^2_\perp$ terms in the
denominators of the propagators \req{propagator}. These terms play 
an important role since they compete with terms $\propto \tau
(\bar{\tau}) Q^2$ which become small in the end-point regions where
either $\tau$ or $\bar{\tau}$ tends to zero. They lead to the 
suppression of contributions with large quark-antiquark separations as
we mentioned above.

In collinear approximation and utilizing \da{}s up to twist-3 accuracy
the subprocess amplitudes for $T\to T$ transitions are infrared
divergent, signaling the break down of factorization \ci{man99}. 
Neglecting transverse momenta in Eq.\ \req{propagator}, one finds 
\be
{\cal H}^V_{++,++}\,+\, {\cal H}^V_{+-,+-}\, \sim
     \int \,\frac{d\tau}{\tau^2\bar{\tau}^2}\, 
     \frac{\xb^2-\xi^2-2\tau\bar{\tau}(\xb^2+\xi^2)}{\xb^2-\xi^2}\,
      \int dk^2_\perp\, k^2_\perp\, \Psi_{VT}(\tau,k^2_\perp)\,.
\label{TT-coll}
\ee
Assuming for instance a Gaussian \wf{} $\Psi_{VT} \sim
\exp{[a_{VT}^2k^2_\perp/(\tau\bar{\tau})]}$, an ansatz that has been
shown to work successfully in many cases (see for instance \ci{jak93})
and will be used by us in the numerical analysis of meson
electroproduction, we find that in fact the $\tau$ integral is
regular. As pointed out in \ci{man99}, the $\xb$ integral in 
Eq.\ \req{amp-nf} may not exist due to the double pole $(\xb-\xi
+i\veps)^{-2}$ occurring. Whether or not this happens depends on 
properties of the GPDs. In Sect.\ \ref{sec:gpd} we will take up this
problem again. 

One may also consider a transverse momentum dependence of
the GPDs. That issue has been investigated in Ref.\ \cite{van99} for
meson electroproduction at intermediate values of $\xbj$. In this
kinematical region the emission and reabsorption of quarks from the
proton dominates. We however think that the $k_\perp$ dependence of
the GPDs is of minor importance. In contrast to the meson where
the hard process enforces the dominance of the compact valence Fock
state of the meson, all proton Fock states contribute to the GPDs at
small $-t$ \cite{DFJK3,diehl03}. If the gluons are distributed in the
proton like the quarks, an assumption that is supported by the
slope of the diffraction peak in elastic proton-proton scattering, 
the $k_\perp$ dependence of the GPD $H^g$ should roughly reflect the
charge radius of the proton ($\langle k^2_\perp \rangle^{1/2}\simeq 200\,\mev$). 
Consequently, we expect $H^g$ to be only mildly dependent on the
transverse momentum, a potential effect we neglect.

%%%%%%%%%%%%%%%%%%%%%%%%%%%%%%%%%%%%%%%%%%%%%%%%%%%%%%%%%%%%%%%%%%%
\section{The impact parameter space}
\label{sec:suda}
%%%%%%%%%%%%%%%%%%%%%%%%%%%%%%%%%%%%%%%%%%%%%%%%%%%%%%%%%%%%%%%%%%%
Transverse momenta in the subprocess amplitudes, see Sect.\ \ref{sec:sub},
imply finite quark-antiquark separations in the configuration space 
which are accompanied by gluon radiation. On the grounds of previous 
work by Collins and Soper \ci{collins:81}, Sterman and collaborators 
\ci{li92} calculated this radiation to next-to-leading-log approximation 
using resummation techniques and having recours to the renormalization 
group. The result is a Sudakov factor which suppresses large 
quark-antiquark separations and which we also have to take into
account in our analysis in order to have consistency with the
retention of the transverse degrees of freedom. Since the Sudakov 
factor is given in the transverse separation or impact parameter space
- only in this space the gluonic radiative corrections exponentiate - 
we have to work in this space. 

The two-dimensional Fourier transformation between the canonical conjugated
$\vbs$ and $\vk$ spaces is defined by 
\be
    \hat{f}(\vbs) \= \frac{1}{(2\pi)^2} \int d^{\,2}\, \vk\, \exp{[-i
                                  \,\vk\cdot\vbs\,]}\; f(\vk)\,.
\label{fourier}
\ee
For the meson \wf{}s we adopt the same Gaussian
parameterization as is used for the pion \cite{jak93,rau96} 
\be
          \Psi_{Vi}(\tau,k^2_\perp)\,=\, 8\pi^2\sqrt{2N_c}\, f_{Vi}\, a^2_{Vi}
       \, \exp{\left[-a^2_{Vi}\, \frac{\vk^{\,2}}{\tau\bar{\tau}}\right]}\,,
\label{wave-l}
\ee
($i=L,T$) which strictly speaking represent full \wf{}s with their
perturbative tails removed. Transverse momentum integration of these
\wf{}s lead to the associated \da{}s which represent the soft hadronic
matrix elements entering calculations within the collinear
factorization approach. Actually, the \wf{} \req{wave-l} lead to the
so-called asymptotic meson \da{} 
\be 
\Phi_{\rm AS} \= 6\tau\bar{\tau}\,.
\label{da-as}
\ee  
For the decay constants, $f_{VL}$ of longitudinally polarized vector
mesons we take the values \ci{stech}:
\be
f_{\rho\, L} \= 0.216\,\gev\,, \qquad  f_{\phi\, L} \= 0.237\,\gev\,.
\label{decay}
\ee
The decay constants for transversely polarized vector mesons are
almost unknown. The only available information comes from QCD sum
rules. In Ref.\ \ci{CZ} $f_{\rho\,T}$ has been estimated to
$(160\pm 10)\,\mev$. We actually fit these decay constants to
experiment. Identifying for instance the soft parameter $M_V$ in the spin
wave function with the meson mass, we obtain 
\be
f_{\rho\, T} \= 0.250\,\gev\,, \qquad f_{\phi\, T} \= 0.275\,\gev\,.
\ee
Choosing $M_V$ to be smaller than the meson mass results in smaller
values of the decay constants $f_{V\,T}$.
The transverse size parameters $a_{VL}$ are fixed by the requirement of
equal probabilities for the vector meson and pion valence Fock states,
namely 0.25. This leads to 
\be
 a_{\rho\, L}\= 0.52\, \gev^{-1}\,, \qquad a_{\phi\, L}\= 0.45\, \gev^{-1}\,.
\label{trans-size}
\ee
The transverse size parameters for transversely polarized vector mesons are
adjusted to experiment. The numerical results we are going to present below
are obtained with
\be
a_{\rho\, T}\= 0.65\, \gev^{-1}\,, \qquad  a_{\phi\, T}\= 0.60\,\gev^{-1}\,.
\ee
With the parameter values quoted in \req{decay} and \req{trans-size}
the r.m.s.\ transverse momenta, evaluated from \req{wave-l}, amount to  
$0.61\,\gev$ and $0.67\,\gev$ for the longitudinally polarized $\rho$
and $\phi$ mesons, respectively.
These values are much larger than the one for the proton r.m.s.\
transverse momentum. 

The Fourier transform of the meson \wf{} (\ref{wave-l}) 
reads
\be
   \hat{\Psi}_{Vi}(\tau,b^2) \= 2\pi\, \frac{f_{Vi}}{\sqrt{2N_c}}\, 
                     \Phi_{\rm AS} (\tau)\, 
               \exp{\left[ -\tau\bar{\tau}\frac{b^2}{4a^2_{Vi}}\right]}\,.
\ee
The product of the propagator denominators $D$ \req{propagator} can be
decomposed into single-pole terms which are either of the form
\be
   {T}_1 \= \frac{1}{\vk^2 + d_1Q^2}\,,
\ee
or
\be
T_2 \= \frac1{\vk^2 - d_2 (\xb-\xi) Q^2-i\hat{\veps}}\,.
\ee
where $d_i\geq 0$. The Fourier transforms of these pole terms can 
readily be obtained
\ba
\hat{T}_1 &=& \frac1{2\pi}\, K_0(\sqrt{d_1}bQ)\,,\nn\\
\hat{T}_2 &=& \frac{1}{2\pi}\, 
           K_0\left(\sqrt{d_2(\xi-\xb)} bQ\right)\;
                    \theta(\xi-\xb) \nn\\
             && + \frac{i}{4} \, 
             H^{(1)}_0\left(\sqrt{d_2(\xb-\xi)}bQ\right)\; 
                        \theta(\xb-\xi)\,,
\ea
where $K_0$ and $H_0^{(1)}$ are the zeroth order modified Bessel
function of second kind and Hankel function, respectively. 

Putting all this together and including the Sudakov factor,
$\exp[-S(\tau,b,Q)]$, the gluonic contributions to the helicity amplitudes for
vector-meson electroproduction read 
\ba
{\cal M}_{\mu'+,\mu+} &=& {\cal M}^H_{\mu'+,\mu+} 
                               + {\cal M}^{\widetilde{H}}_{\mu'+,\mu+}\,,\nn\\[0.3em]
{\cal M}^H_{\mu'+,\mu+} &=& \frac{e}{\sqrt{2N_c}}\, {\cal C}_V \;
                    \int d\xb d\tau\, \,f^+_{\mu'\mu}\,  H^g(\xb,\xi,t)\nn\\[0.3em]
            &\times& \int d^{\,2} \vbs\, \hat{\Psi}_{V\mu^\prime} (\tau,b^2)\,
                           \hat{D} (\tau,Q,b)\,\als(\mu_R)\,  
                           \exp{[-S(\tau,b,Q)]}\,,\nn\\[0.3em] 
{\cal M}^{\widetilde{H}} _{\mu'+,\mu+} &=& \frac{e}{\sqrt{2N_c}}\, {\cal C}_V \;
          \int d\xb d\tau \,f^-_{\mu'\mu}\, \widetilde{H}^g(x,\xi,t)\,\nn\\[0.3em]
      &\times& \int d^{\,2} \vbs\, \hat{\Psi}_{V\mu^\prime} (\tau,b^2)\,
                           \hat{D} (\tau,Q,b)\,\als(\mu_R)\,  
                           \exp{[-S(\tau,b,Q)]}\,, 
\label{mod-gluon}
\ea
which is the $\vbs$-space version of the amplitude (\ref{amp-nf}). The
functions $D$ and $f^\pm_{\mu'\mu}$ are given in \req{propagator} and
\req{hard-amp-ji}. Since the Fourier transformed \wf{}s, the 
product of propagator denominators as well as the Sudakov factor only
depend on $b$ the angle integration in the last integral is trivial
and a three-dimensional integral ($d\xb d\tau b\,db$) remains to be
evaluated numerically. Parity invariance \req{parity} lead to the
following relations among the amplitudes~\footnote{The same relations
  as for the $H^g$ terms also hold for the $t$-channel exchange of a
  particle with natural parity, $P=(-1)^J$. The relations for the
  $\widetilde{H}^g$ terms in \req{mod-gluon} corresponds to those
  obtained for an unnatural parity exchange \ci{schi}.} 
\ba
{\cal M}^H_{++,++} &=& \phantom{-}{\cal M}^H_{-+,-+}\,, \qquad
             {\cal M}^H_{0+,++} \= -{\cal M}^H_{0+,-+} \,,\nn\\
{\cal M}^{\widetilde{H}}_{++,++} &=& -{\cal M}^{\widetilde{H}}_{-+,-+}\,, \qquad
   {\cal M}^{\widetilde{H}}_{0+,++} \=\phantom{-} {\cal M}^{\widetilde{H}}_{0+,-+} \,, 
\label{npe}
\ea

The Sudakov exponent $S$ in (\ref{mod-gluon}) is given by \ci{li92}
\be
S(\tau,b,Q)=s(\tau,b,Q)+s(\bar{\tau},b,Q)-\frac{4}{\beta_0}
\ln\frac{\ln\left(\mu_R/\Lambda_{QCD}\right)}
                                {\hat{b}}\,,
\label{sudeq}
\ee
where a Sudakov function $s$ occurs for each quark line entering the
meson and the abbreviation 
\be
  \hat{b} \= -\ln{\left(b\LQCD\right)} \,,
\ee
is used. The last term in (\ref{sudeq}) arises from the application of the
renormalization group equation ($\beta_0=11-\frac{2}{3} n_f$) where
$n_f$ is the number of active flavors taken to be 3. A value
of $220$ MeV for $\Lambda_{QCD}$ is used here and in the evaluation of
$\als$ from the one-loop expression. The renormalization scale $\mu_R$
is taken to be the largest mass scale appearing in the hard scattering 
amplitude, i.e. $\mu_R=\max\left(\tau Q,\bar{\tau}Q,1/b\right)$. For
small $b$ there is no suppression 
from the Sudakov factor; as $b$ increases the Sudakov factor decreases,
reaching zero at $b=1/\Lambda_{QCD}$. For even larger $b$ the Sudakov is set 
to zero~\footnote
{The definition of the Sudakov factor is completed by the following
rules \cite{li92}: $\exp{[-S]}=1$ if $\exp{[-S]}\geq 1$,
$\exp{[-S]}=0$ if $b\geq 1/\LQCD$ and $s(\beta,b,Q)=0$ if $b\leq
\sqrt{2}/\beta Q$.}. 
The Sudakov function $s$ reads
\be
 s(\tau,b,Q) = \frac{8}{3\beta_0} \left( \hat{q} 
                       \ln{\left(\frac{\hat{q}}{\hat{b}}\right)}
                       -\hat{q} + \hat{b} \right) + {\rm NLL-terms}\,,
\ee
where
\be
     \hat{q} \= \ln{\left(\tau Q/(\sqrt{2}\LQCD)\right)}\,.
\ee
Actually we do not use the explicit form of the next-to-leading-log
corrections quoted in \ci{li92} but those given in Ref.\ \cite{DaJaKro:95}. 
The latter ones contain some minor corrections which are
hardly relevant numerically. Due to the properties of the Sudakov
factor any contribution to the amplitudes is damped asymptotically,
i.e. for $\ln (Q^2/\LQCD^2)\to\infty$, except those from
configurations with small quark-antiquark separations. $b$ plays the 
role of an infrared cut-off; it sets up the interface between 
non-perturbative soft gluon contributions - still contained in the 
hadronic \wf{} - and perturbative soft gluon contributions accounted 
for by the Sudakov factor. 

%%%%%%%%%%%%%%%%%%%%%%%%%%%%%%%%%%%%%%%%%%%%%%%%%%%%%%%%%%%%%%%%%%%%
\section{Modelling the GPDs}
\label{sec:gpd}
%%%%%%%%%%%%%%%%%%%%%%%%%%%%%%%%%%%%%%%%%%%%%%%%%%%%%%%%%%%%%%%%%%%
In order to calculate the electroproduction amplitudes \req{amp-nf} we
still need the gluon GPDs. A model for them can be constructed 
with the help of double distributions \ci{mus99} which guarantee 
polynomiality of the GPDs. The gluonic double distribution 
$f(\beta,\alpha,t\simeq 0)$ is customarily parameterized as 
\be
f(\beta,\alpha,t\simeq 0) \= g(\beta)\,
                 \frac{\Gamma(2n+2)}{2^{2n+1}\,\Gamma^2(n+1)}\,
               \frac{[(1-|\beta|)^2-\alpha^2]^n}{(1-|\beta|)^{2n+1}} \,,
\label{double-dis}
\ee
where $g(x)$ is the usual gluon distribution. Its definition is
extended to negative $\beta$ by 
\be
g(-\beta) \= -g(\beta)\,.
\ee
A popular choice of $n$ is 1 for quarks and $2$ for gluons. This is
motivated by the interpretation of the $\alpha$ dependence like a
meson \da{} for hard exclusive processes. The cases $n=1$ and 2
correspond to the asymptotic behavior of a quark \da{} 
$\propto (1-\alpha^2)$ and for gluons $\propto (1-\alpha^2)^2$,
respectively. This correspondence is not demanded by
theory. Therefore, we will consider both the cases, $n=1$ and 2, for
the construction of the gluon GPD. A parameterization of the $t$ dependence of
$f$ is difficult. The multiplication of $f$ as given in \req{double-dis}  
by a $t$-dependent form factor, although frequently used in default of a
better idea, is unsatisfactory. Parameterizations of the GPDs
\ci{DFJK4,DFJK1} as well as results from lattice QCD \ci{haegler}
revealed that a factorization of $f$ in $\beta,\alpha$ on the one hand
and in $t$ on the other hand is most likely incorrect. Fortunately,
the knowledge of the GPDs at $t\simeq 0$ suffices for our purposes as
will become clear in Sect.\ \ref{sec:cross}.

According to \ci{mus99}, the GPD $H^g$ is related to the double
distribution by (since we will only work with GPDs at $t\simeq 0$ we
omit the variable $t$ in the GPDs in the following)
\ba
H^{g}(\xb,\xi) &=&  \Big[\,\Theta(0\leq \xb\leq \xi)
         \int_{x_3}^{x_1}\, d\beta + 
       \Theta(\xi\le \xb\leq 1) \int_{x_2}^{x_1}\, d\beta \,\Big] 
        \frac{\beta}{\xi}\,f(\beta,\alpha=\frac{\xb-\beta}{\xi})\nn\\
        &&         + \xi\, D^g\Big(\frac{\xb}{\xi}\Big)\,. 
\label{gpd-model}
\ea
The definition of $H^g$ is completed by noting that it is an even
function of  $\xb$ 
\be
    H^g(-\xb,\xi) \= H^g(\xb,\xi)\,.
\ee
The integration limits in Eq.\ \req{gpd-model} are given by
\be
x_1\=\frac{\xb+\xi}{1+\xi}\,,  \quad x_2\=\frac{\xb-\xi}{1-\xi}\,,
\quad x_3\= \frac{\xb-\xi}{1+\xi}\,.
\ee
The limit  $x_1 (x_2)$ is the momentum fraction the emitted (reabsorbed)
gluon carries with respect to the incoming (outgoing) proton.
The last term in the definition \req{gpd-model} is the so-called
$D$-term~\ci{pol99}. Its support is the region $-\xi\leq \xb\leq \xi$
and it ensures the correct polynomiality property of the GPD. Since
the $D$ term is $\propto \xi$ and our interest lies in small skewness,   
we neglect it.

We take the gluon distribution from the NLO
CTEQ5M results \ci{CTEQ} and use an interpolation of it which has been
proposed in Ref.\ \ci{MRST} and which is valid
in the  range $Q_0^2=4\,\gev^2 \leq Q^2 \leq 40\, \gev^2$  
\be
\beta g(\beta)\= \beta^{-\delta(Q^2)}\,(1-\beta)^5\,
                 \sum_{i=0}^2 c_i\, \beta^{\,i/2}\,,
\label{dg}
\ee 
where
\be
c_0\= 1.94\,,\quad c_1\= -3.78 + 0.24 \ln{({Q^2}/{Q^2_0})}\,,
      \quad c_2\= 6.79-2.13\ln{({Q^2}/{Q^2_0})} \,.
\label{dg-coeff}
\ee
This is a very good approximation to the CTEQ gluon distribution for
$\beta \leq 0.5$. At the largest value of $Q^2$ we are going to use the
interpolation \req{dg}, namely $40 \,\gev^2$, and for $\beta \simeq
0.2$ it deviates less than $5\%$ from the CTEQ gluon distribution. For
values of $\beta$ in the 
range $10^{-4}- 10^{-1}$, the interpolation \req{dg} agrees
with the CTEQ gluon distribution within $1\%$. In this region 
the $Q^2$-dependence of $\delta$ is approximatively given by 
\be
\delta(Q^2) \= 0.17 + 0.07 \ln{(Q^2/Q_0^2)} - 0.005 \ln^2{(Q^2/Q_0^2)}\,.
\label{kappa}
\ee
The parameterization \req{dg}, \req{dg-coeff} and \req{kappa} effectively
take into account the evolution of the gluon distribution in a
large but finite range of $Q^2$ as calculated in Ref.\ \ci{CTEQ}.  
At small $\beta$ the gluon distribution has a typical error of about
$15\%$ \ci{CTEQ}. Within this error there is agreement with the
analysis presented in Ref.\ \ci{MRST}. An error assessment of the power
$\delta$ provides an uncertainty of about $10-15\%$ for it \ci{CTEQ,MRST}. 

For the various terms in the ansatz \req{dg} the integrations occurring
in Eq.\ \req{gpd-model} can be performed analytically, see Ref.\
\ci{mus99}. One finds
\ba
H_{1i}(\xb,\xi)&=& \frac{3}{2\xi^3}\,
     \frac{\Gamma(1+i/2-\delta)}{\Gamma(4+i/2-\delta)}\;\nn\\
    &\times& \Big\{ \,(\xi^2-\xb)\, \Big[x_1^{2+i/2-\delta}-x_2^{2+i/2-\delta}\Big] \nn\\
    &&\qquad +\; \xi(1-\xb)(2+i/2-\delta)\,\Big[x_1^{2+i/2-\delta}+ 
     x_2^{2+i/2-\delta}\Big]\, \Big\} \,, \qquad \xb\geq\xi\,, \nn\\
               &=& \frac{3}{2\xi^3}\, 
                   \frac{\Gamma(1+i/2-\delta)}{\Gamma(4+i/2-\delta)}\; 
      \Big\{\,x_1^{2+i/2-\delta}\,\Big[\xi^2-\xb +(2+i/2-\delta)\xi (1-\xb)\Big] \nn\\
              && \qquad +\; (\xb\to -\xb)\,\Big\}\,, \hspace{5.6cm} \xb\leq\xi\,, 
\label{H-model1}
\ea
for the case $n=1$. A similar but somewhat more complicated
results is obtained for the case $n=2$.
%\ba
%H_{2i}(\xb,\xi)&=& \frac{15c_i}{4\xi^5}\,
%                   \frac{\Gamma(1+i/2-\delta)}{\Gamma(6+i/2-\delta)}\,\nn\\[0.2em]
%         &\times& \Big\{\,\Big[\big[3(3+i/2-\delta)^2 
%           -(1+i/2-k)(5+i/2-k)\big]\,(\xi^2-\xb)^2\,  \nn\\[0.2em]
%        &&  -2(2+i/2-\delta)(4+i/2-\delta)(1-\xi^2)(\xb^2-\xi^2)\Big]\,       
%            \Big[x_1^{3+i/2-\delta}-x_2^{3+i/2-\delta}\Big] \nn\\[0.2em]
%     &&+\; 6(3+i/2-k)\,\xi(1-\xb)(\xi^2-\xb)\,
%                    \Big[x_1^{3+i/2-\delta}+x_2^{3+i/2-\delta}\Big]\,\Big\}\,,
%                            \quad \qquad \xb\geq\xi\,, \nn\\[0.2em]
%            &=&\frac{15c_i}{4\xi^5}\,\frac{\Gamma(1+i/2-\delta)}
%               {\Gamma(6+i/2-\delta)}\,\Big\{ \,x_1^{3+i/2-\delta}\,
%                  \Big[\big[3(3+i/2-\delta)^2\nn\\[0.2em]
%           &&-(1+i/2-k)(5+i/2-k)\big](\xi^2-\xb)^2  
%                    +\;6(3+i/2-\delta)\xi(\xi^2-\xb)(1-\xb) \nn\\[0.2em]
%          &&-2(2+i/2-\delta)(4+i/2-\delta)(1-\xi^2)(\xb^2-\xi^2)\Big] \nn\\[0.2em]  
%            &&+\;(\xb\to -\xb)\,\Big\}\,, \hspace{7.7cm} \xb\leq\xi\,. 
%\label{H-model2}
%\ea
This way we obtain an expansion of $H^g$
\be
H^g(\xb,\xi) \= \sum_i \hat{c}_{ni}\, H_{ni}(\xb,\xi)\,,
\label{H-sum}
\ee 
with coefficients following from \req{dg}. The evolution of the GPD
is here approximated by that of the gluon distribution. The dominant 
contribution to vector meson electroproduction comes from the
imaginary part of the $L\to L$ amplitude (see Sect.\ \ref{sec:cross}) 
which is $\propto H^g(\xi,\xi)$. Since, to a good approximation, 
$H^g(\xi,\xi)$ equals $\xbj\, g(\xbj)$ at small $\xi$ we have 
approximately taken into account the evolution. The 
full evolution of the gluonic GPD is complicated because of mixing 
with the flavor-singlet quark GPD. Its modelling would counteract any
possible gain of accuracy obtained by the inclusion of the full
evolution behavior.

\begin{figure}[t]
\begin{center}
\includegraphics[width=.48\textwidth,bb=63 355 530 747,clip=true]{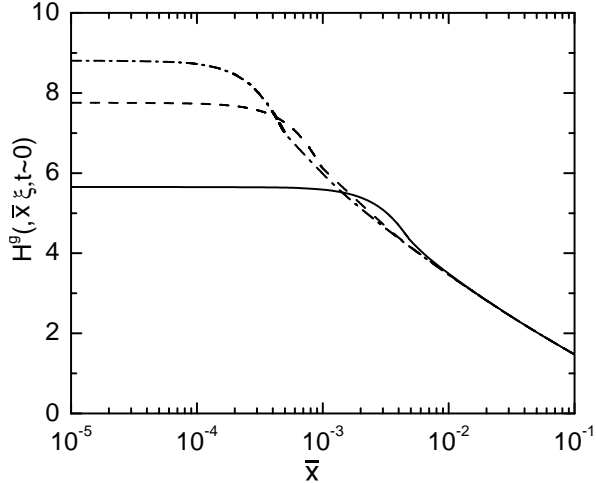}
\end{center}
\caption{Model results for the GPD $H^g$ in the small $\xb$ range 
 at $t\simeq 0$ and for the case $n=1$. 
The solid (dashed,  dash-dotted) line represents the GPD at  
$\xi= 5\;(1\,,\; 0.5)\, \cdot 10^{-3}$ and at the scale $2\,\gev$.}
\label{fig:H}
\end{figure}

The GPD $H^g$ and its derivatives up to order $n$ are continuous at
$\xb=\xi$. For $\xi\ll \xb$ one can convince oneself that
$H^g(\xb,\xi)\to \xb\,g(\xb)$ up to corrections of order $\xi^2$.
In the forward limit, $\xi, t \to 0$, the GPD $H^g$ reduces to the 
ordinary parton distribution $\xb g(\xb)$. Results for $H^g$ in the 
case $n=1$ are shown in Fig.\ \ref{fig:H}. For $\xb$ 
larger than $\xi\ll 1$  there is practically no dependence on the
skewness in contrast to the region $\xb\leq \xi$ in accord with the 
general behavior of the model GPD just mentioned. The GPDs for $n=1$
and 2 agree with each other on the percent level at small $\xb$. As we
checked the numerical results for the cross sections obtained  
with both these GPDs are very similar; the differences in the
imaginary (real) parts of the amplitudes are typically smaller than 
$1 (7)\%$. In the following we will therefore show only numerical 
results for the case $n=1$.

Considering the collinear 
limit of the subprocess amplitude \req{TT-coll}, one notices a double 
pole $(\xb-\xi+i\veps)^{-2}$ occurring in the $T\to T$ amplitude 
\req{amp-nf} \ci{man99}. Partial integration leads to the integral
\be
\sim \int_0^1 \frac{d\xb}{\xb-\xi+i\veps}\,
              \frac{d}{d\xb}\Big[H^g(\xb,\xi)\tilde{f}(\xb,\xi)\Big]\,,
\ee
where $\tilde{f}$ arises from the subprocess amplitude
\req{TT-coll}. Since the derivatives of $H^g$ and $\tilde{f}$ are 
continuous at $\xb=\xi$ the integral exist. The transverse quark
momenta are not needed for the regularization of the $T\to T$ amplitude.

A model for the GPD $\widetilde{H}^g$ can be constructed analogously 
to \req{double-dis}, \req{gpd-model}, the parton distribution $g(\beta)$ 
is only to be replaced by its polarized counterpart $\Delta g(\beta)$.
The continuation to negative $\beta$ is defined by
\be 
\Delta g(-\beta)\=\Delta g(\beta)\,.
\ee  
The GPD $\widetilde{H}^g$ is antisymmetric in $\xb$
\be
\widetilde{H}^g(-\xb,\xi) = - \widetilde{H}^g(\xb,\xi)\,.
\ee 
We take $\Delta g$ from Ref.\ \ci{BB} and parameterize it analogously
to \req{dg}
\be
\beta \Delta g(\beta)\= \beta^{\tilde{\delta}(Q^2)}\, (1-\beta)^5\, 
                  \sum_{i=0}^{2}\, \tilde{c}_i\, \beta^i\,,
\label{eq:glu-pd}
\ee
where
\ba
\tilde{c}_0 &=& 3.39-0.864\, \ln(Q^2/Q^2_0)\,, \quad \tilde{c}_1\=
1.73 + 0.24\, \ln(Q^2/Q^2_0) -0.17\,\ln^2(Q^2/Q^2_0) \nn\\ 
\tilde{c}_2 &=& 0.42 -0.115\,\ln(Q^2/Q^2_0)
-0.069\,\ln^2(Q^2/Q^2_0)\,,
\ea
and
\be
\tilde{\delta}(Q^2) \= 0.78 -0.173\, \ln(Q^2/Q^2_0)\,.
\ee   
The GPD $\widetilde{H}^q$ can than calculated analytically for either
case, $n=1$ and 2, with, for instance, the help of \req{H-model1}. It
is then represented by a sum analogously to \req{H-sum}. We finally
remark that the polarized gluon distribution and hence
$\widetilde{H}^g$ is subject to much larger uncertainties than $H^g$. 

%%%%%%%%%%%%%%%%%%%%%%%%%%%%%%%%%%%%%%%%%%%%%%%%%%%%%%%%%%%%%%%%%%%%%%%%%
\section{Cross sections}
\label{sec:cross}
%%%%%%%%%%%%%%%%%%%%%%%%%%%%%%%%%%%%%%%%%%%%%%%%%%%%%%%%%%%%%%%%%%%%%%%%%
Vector meson electroproduction in the diffractive region has been
extensively investigated at HERA \ci{h1,zeus98,zeus99,adloff,zeus96,h196} 
for large $W$ and $Q^2$ but small $\xbj$. Preliminary data
from H1 and ZEUS \ci{h1-prel,ZEUS-prel-rho,ZEUS-prel-phi} extend
the range of $Q^2$ for which electroproduction data are available. In order
to confront the data with the theory developed in the preceding
sections, one has either to extrapolate the data to $t\simeq 0$ or to
take into account the $t$ dependencies of the GPD and the subprocess
amplitudes. The latter recipe is not straightforward. As we mentioned 
in Sect.\ \ref{sec:gpd} it is not easy to find a plausible
parameterization for the $t$ dependence of the GPD because
factorization in $t$ and $\xb, \xi$ most likely does not hold \ci{DFJK1,haegler}. 
We therefore use a variant of the first recipe and multiply the 
$t\simeq 0$ amplitudes \req{amp-nf}, \req{mod-gluon} by exponentials
\be
        \sim \exp{[t\, B_i^V/2]}\,,
\label{slope}
\ee
with slope parameters, $B^V_i$ ($i=LL, LT, TT$ for $L\to L, T\to L,
T\to T$ transitions, respectively), adjusted to experiment. The
ansatz \req{slope} is in accord with the expected exponential
behavior of the GPDs \ci{DFJK4,goeke}. Differences in
the slope parameters arise from the t dependence of the subprocess
amplitudes. 

In the one-photon exchange approximation the $ep\to epV$ cross section
integrated over the azimuthal angle , reads 
\be
\frac{d^3 \sigma(ep\to ep\rho)}{dy dQ^2 dt} 
       \= \frac{\ale}{2\pi}\,\frac{1+(1-y)^2}{y\,Q^2}\,
       \left[\,\frac{d\sigma_T}{dt} {+} \veps \frac{d\sigma_L}{dt}\,\right]\,,
\label{epcross}
\ee 
where high-energy, small $\xbj$ approximations have been applied
to the phase space factor. Under the same kinematical conditions the
ratio of longitudinal to transversal polarization of the virtual
photon is given by
\be
\veps \simeq  2\frac{1-y}{1 {+} (1-y)^2}\,,
\label{pho-pol}
\ee
where $y$ is the fraction of longitudinal electron momentum carried
by the photon
\be
y \= \frac{q\cdot p}{k_e\cdot p}\= \frac{W^2+Q^2}{s}\,.
\ee
Here, $k_e$ is the momentum of the incident electron and $s=(k_e+p)^2$. The 
$\gamma^* p\to Vp$ partial cross sections in \req{epcross} 
for transversally and longitudinally polarized virtual photons are
related to the amplitudes \req{mod-gluon} by
\ba
\frac{d\sigma_T}{dt}&=&\frac{1}{16\pi W^2\,(W^2 {+} Q^2)}\,\Big[\,\big|{\cal
      M}_{++,++}^H\big|^2 {+} \big[{\cal M}^H_{0+,++}\big|^2\,\Big]\,,\nn\\
\frac{d\sigma_L}{dt}&=&\frac{1}{16\pi W^2 \,(W^2+Q^2)}\, 
                \big|{\cal M}_{0+,0+}^H\big|^2 \,,
\label{cross-lt}
\end{eqnarray}
where we made use of Eqs.\  \req{parity}, \req{hierarchy} and
\req{npe}. Terms of order $\langle\widetilde{H}^g\rangle^2$ have been
neglected in the cross sections \req{cross-lt}; there 
is no interference between the $H^g$ and $\widetilde{H}^g$ contributions.

The differential cross section data for $ep\to epV$ exhibit a characteristic 
diffraction peak at small $t$. The slope of the diffraction peak is 
found to be nearly independent of $W$ but is mildly varying with $Q^2$. 
Most of the differential cross section data for $\rho$ and $\phi$ production
are compatible with a single exponential within errors. The combined
H1 and ZEUS data on the slopes in the range $4\,\gev^2 \lsim Q^2 \lsim
40 \gev^2$ can be condensed into
\be   
B^{\,V}_{LL} \= 7.5\,\gev^{-2} + 1.2\,\gev^{-2}
                                \ln{\frac{3.0\,\gev^2}{Q^2+ m^2_V}}\,.
\label{bslope}
\ee
This parameterization is in rather good agreement with experiment,
keeping in mind that the experimental slopes are not always extracted 
from cross section data in the same range of $t$. Possible deviations
from a single exponential behavior of the cross sections then
lead to different slopes. We naturally assign the slope \req{bslope}
to the dominant $L\to L$ transition amplitude. The slopes of the other
amplitudes are not well determined as yet. A detailed analysis of
the spin density matrix elements presented in Sect.\ \ref{sec:sdme},
seem to favor the choice $B^V_{LT}=2 B^V_{TT}=B^V_{LL}$
slightly. These slope values lead to results from our GPD based
approach in fair agreement with the HERA data. It is to be stressed
that the magnitude of the transverse cross section is controlled by
the product of parameters $(f_{V\,T}/M_V)^2/B^V_{TT}$ leaving aside the mild $Q^2$
dependence of the slope. The just described fit is based on the choice
$M_V=m_V$ (see the remark subsequent to Eq.\ \req{spin01}). Taking a
smaller value for $M_V$ and a corresponding value for the decay
constant, the slope $B^V_{TT}$ can be closer to that one for the $L\to
L$ amplitude. For instance, choosing $M_V=m_V/2$, one may use
$B^V_{TT}=B^V_{LL}$ ( for $f_{\rho\,T}=170\,\mev$ and
$f_{\phi\,T}=190\,\mev$) and obtaines almost identical results for
the cross sections. 

\begin{figure}[t]
\begin{center}
\includegraphics[width=.48\textwidth,bb=25 335 520 732,clip=true]{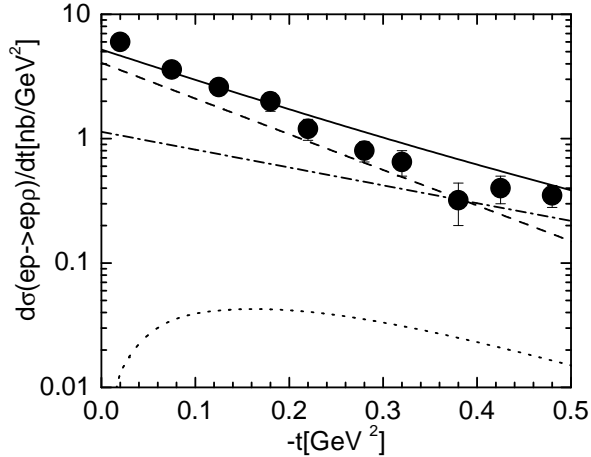}
\end{center}
\caption{The differential cross section \req{epcross} for $ep\to ep\rho$
  versus $-t$ integrated over the kinematical region available to the ZEUS
  experiment from which the data are taken \protect\ci{zeus98}. The
  solid line is our fit to the data at $W = 75\,\gev$ and
  $Q^2 = 6\,\gev^2$ (see text). The dashed (dot-dashed, 
  short-dashed) line represents the individual contributions from the 
  $L\to L$ ($T\to T$, $T\to L$) amplitudes.}
\label{fig:dsdt}
\end{figure}
As a check of our choice of the slopes we show the ZEUS
data~\ci{zeus98} for the differential cross section of $\rho$
production in Fig.\ \ref{fig:dsdt}. These data indicate deviations 
from a single exponential behavior. They are integrated over the 
$W$ and $Q^2$ region accessible to ZEUS; $W$ varies between 32 and 
$167\,\gev$ in dependence on $Q^2$ which varies between 3 and 
$50\,\gev^2$. The associated normalization uncertainty 
is of no bearing to us since we are interested in the process 
$\gamma^* p\to Vp$. The forward amplitudes \req{mod-gluon} evaluated 
from the model GPD $H^g$ shown in Fig.\ \ref{fig:H}, multiplied with the 
exponentials \req{slope}, lead to the results for the $ep\to ep \rho$ 
differential cross section shown in Fig.\ \ref{fig:dsdt}. The
agreement between our result and experiment is not too good. Obviously, 
the value of the slope taken from Eq.\ \req{bslope} at $Q^2=6\,\gev^2$, 
is a bit too small. However, the data shown in Fig.\ \ref{fig:dsdt},
need confirmation. We can also see from the figure that our result
although obtained with
different slopes, do not deviate from a single exponential behavior 
substantially.  Also shown in Fig.\ \ref{fig:dsdt} are the three 
individual contributions $L\to L$, $T\to T$ and $T\to L$ separately.
As expected the $L\to L$ contribution dominates. The $T\to T$
contributions amounts to about $25\%$ of the $L\to L$ one at $t\simeq
0$. Due to the smaller slope $B^V_{TT}$ takes the lead for $-t$ larger
than about $0.4\,\gev^2$. The $T\to L$ contribution is shown only for
comparison, it is of no account to the cross sections.

\begin{figure}[t]
\begin{center}
\includegraphics[width=.44\textwidth,bb=36 348 540
  741,clip=true]{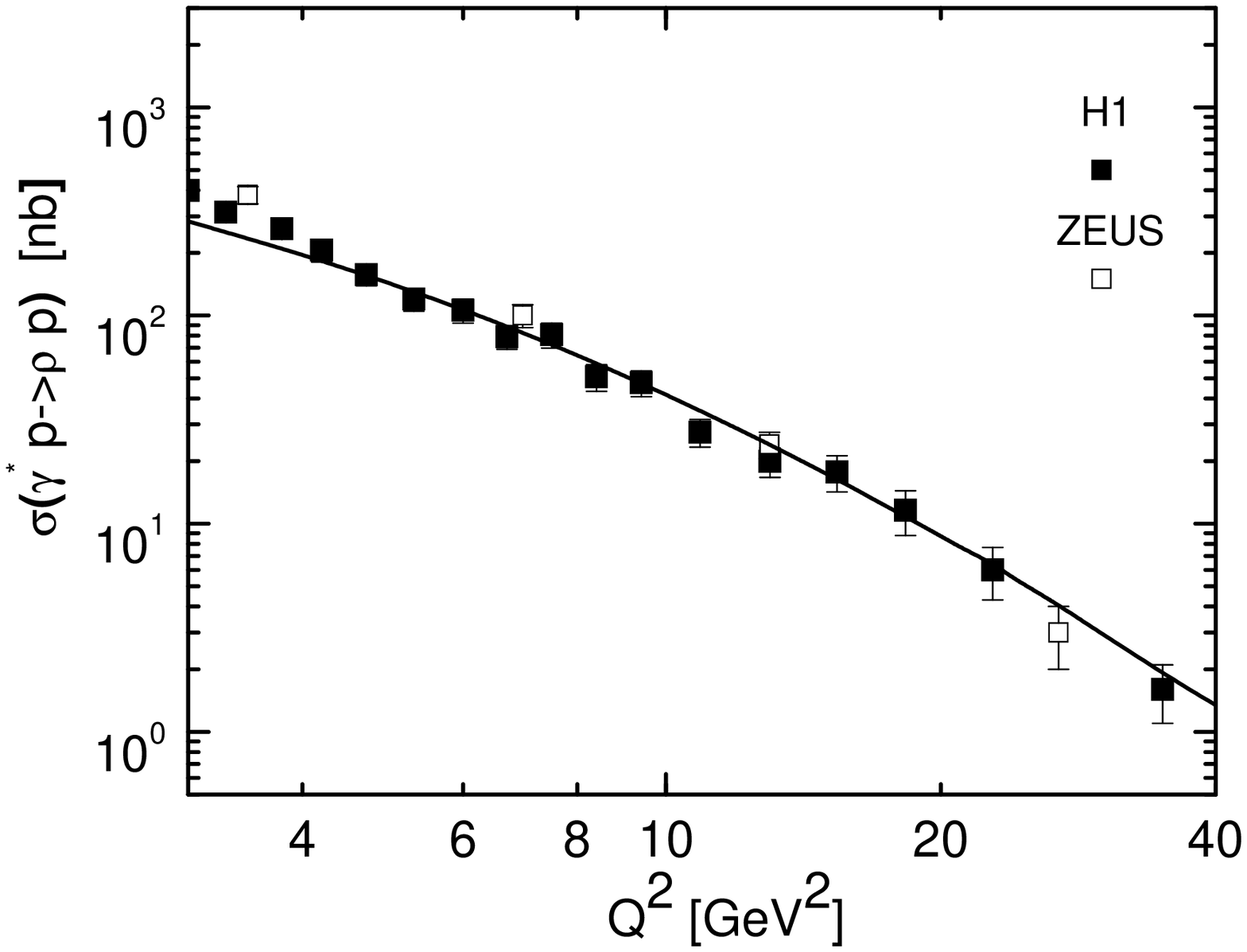}\hspace*{0.3cm} 
\includegraphics[width=.44\textwidth,bb=29 336 555 741,clip=true]{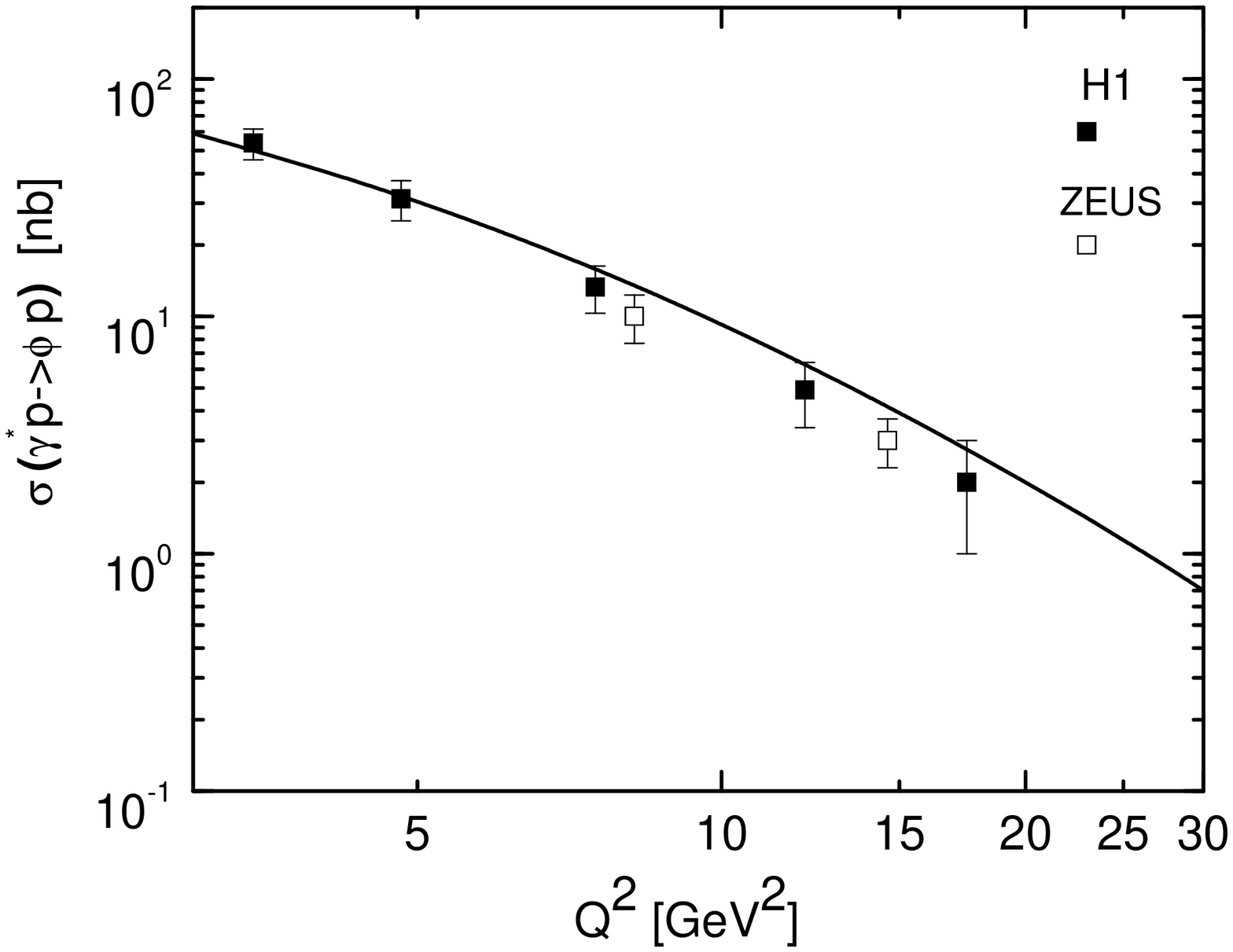}
\end{center}
\caption{The integrated cross section for $\gamma^* p\to \rho p$
  (left) and $\gamma^* p\to \phi p$ (right) versus $Q^2$ at 
  $W\simeq 75\, \gev$. Data taken from \protect\ci{h1,adloff} (filled
  squares) and \protect\ci{zeus98,zeus96} (open squares) for $\rho$ and 
  $\phi$ production, respectively. The solid lines represent our results.}
\label{fig:sigrho}
\end{figure}

Let us now turn to the discussion of the process $\gamma^* p \to Vp$. 
The integrated cross section for this process 
is related to the integrated partial cross sections \req{cross-lt} by
\be 
\sigma(\gamma^*p\to Vp) \= \sigma_T(\gamma^*p\to Vp) + 
                              \eps\, \sigma_L(\gamma^*p\to Vp)\,.
\label{int-cross}
\ee 
The H1~\ci{h1,adloff} and ZEUS~\ci{zeus98,zeus96} data on the cross sections
for $\gamma^* p\to pV$ ($V=\rho,\phi$), integrated over the
diffraction peak, are compared to our results in Fig.\
\ref{fig:sigrho}. We repeat our results are evaluated from the handbag
amplitude \req{mod-gluon} multiplied by the exponentials \req{slope}
and using the GPD $H^g$ shown in Fig.\ \ref{fig:H}. 
Good agreement between model and experiment is achieved for both
processes provided $Q^2$ is larger than about $4\,\gev^2$.

\begin{figure}[pt]
\begin{center}
\includegraphics[width=.44\textwidth,bb=38 343 540 741,clip=true]
{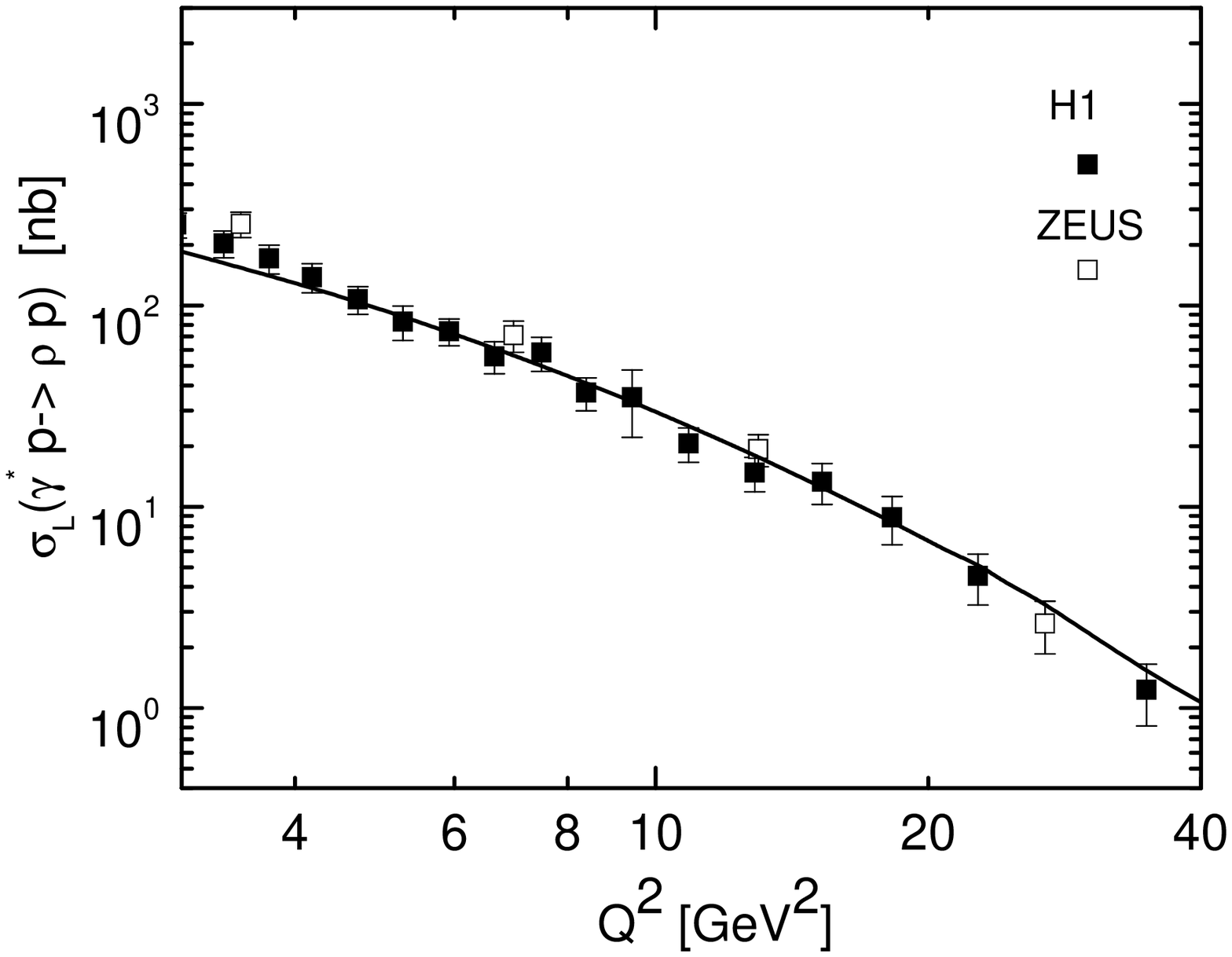}\hspace*{0.3cm}
\includegraphics[width=.44\textwidth,bb=27 327 555 741,clip=true]{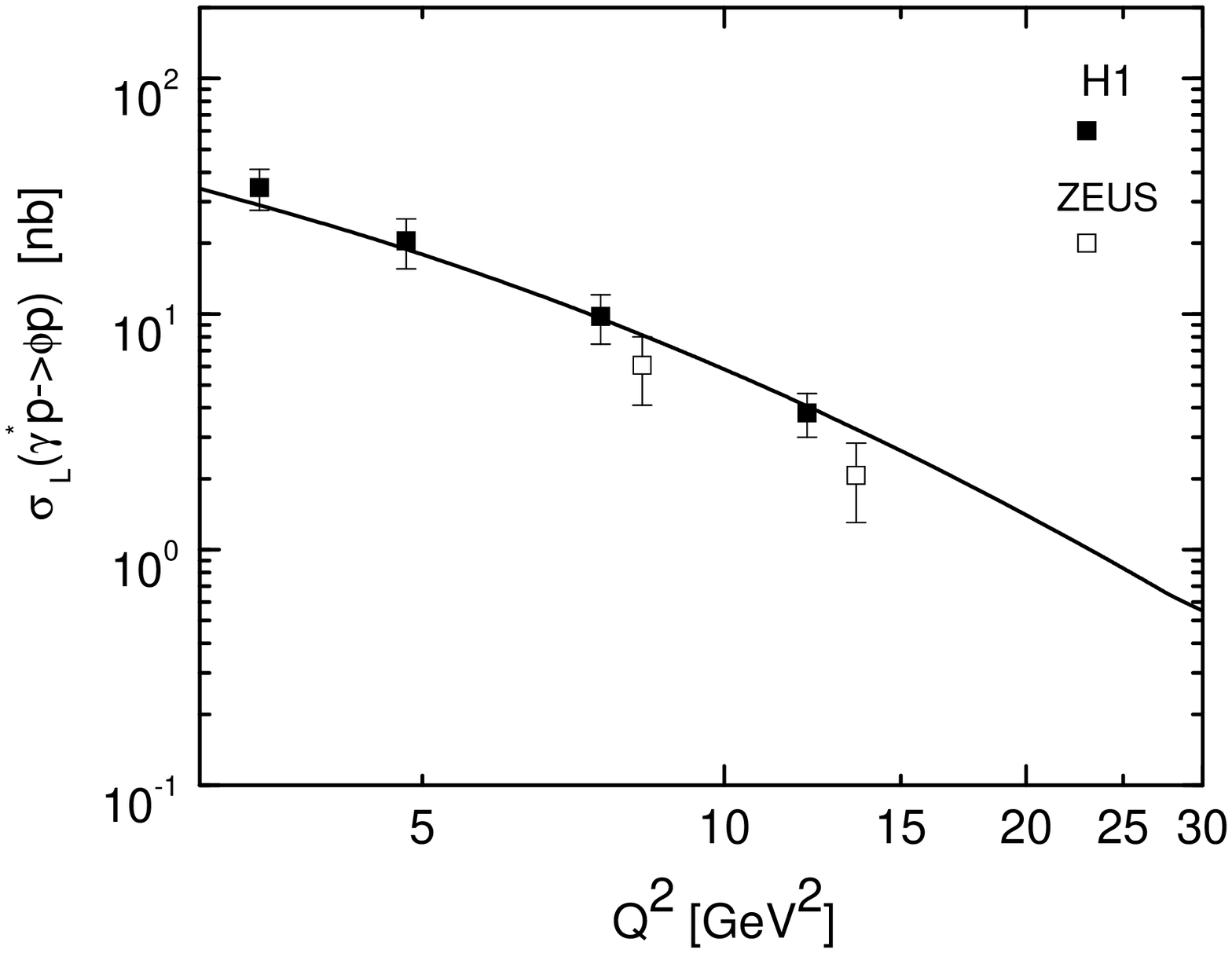}
\end{center}
\caption{The integrated cross section for longitudinally polarized
  photons versus $Q^2$ at $W\simeq 75\, \gev$. Left: 
  $\gamma^*_L\, p\to \rho p$; right: $\gamma^*_L\, p\to
  \phi p$. The data are taken from H1 \protect\ci{h1,adloff,h196} 
  (filled squares) and ZEUS \protect\ci{zeus98,zeus96} (open squares),
  respectively. The solid lines represent our results.} 
\label{fig:sigrhoL}
%\end{figure}
%\nopagebreak
%\begin{figure}[h]
\begin{center}
\includegraphics[width=0.44\textwidth, bb= 49 352 530 733,clip=true]
{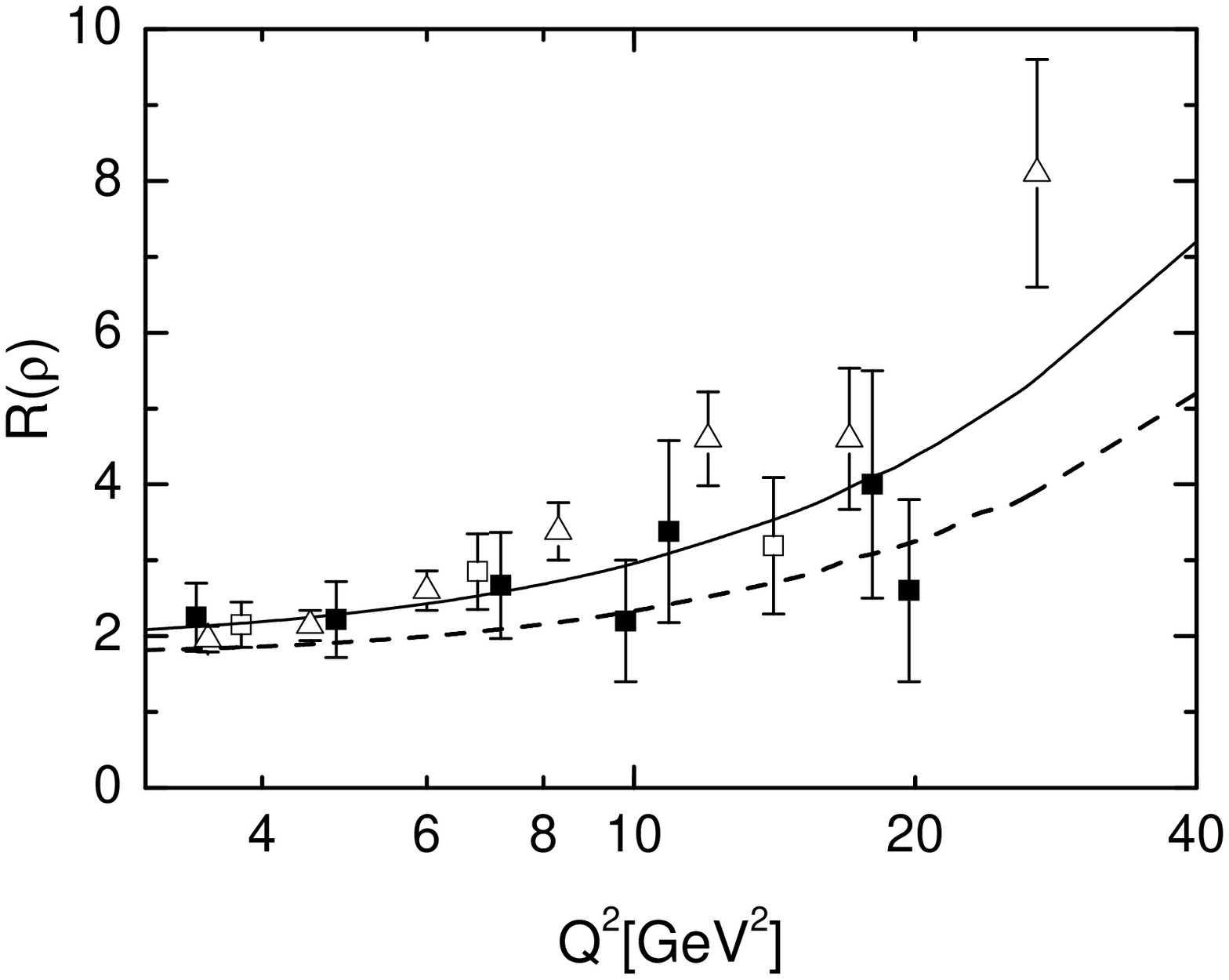}\hspace*{0.3cm}
\includegraphics[width=0.44\textwidth, bb= 44 346 528 733,clip=true]{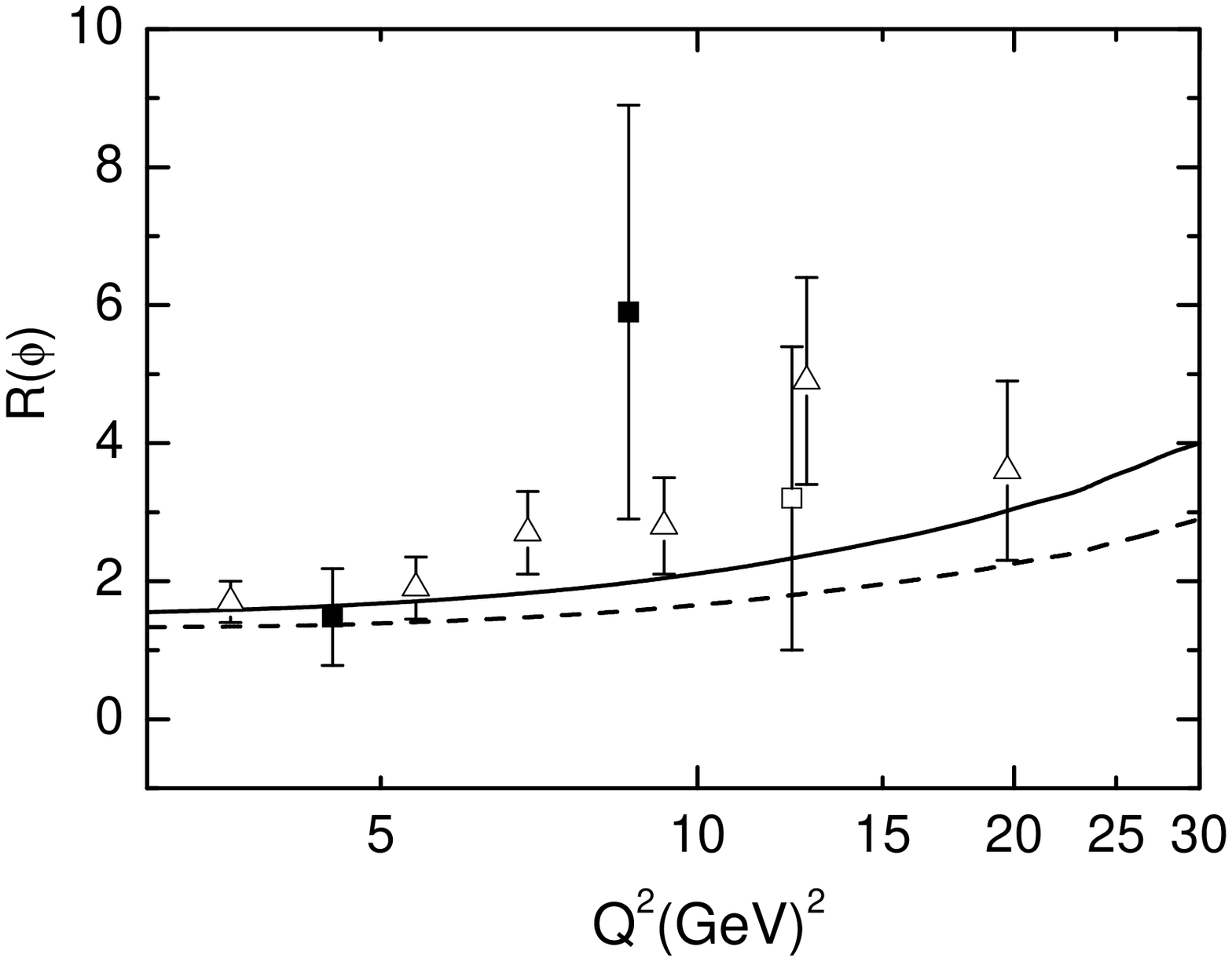}
\caption{The ratio of longitudinal and transverse cross sections
 for $\rho$ (left) and $\phi$ (right) production versus $Q^2$ at
 $W \simeq 75\, \gev$. Data taken from
 \protect\ci{h1,adloff,h196} (filled squares) and \protect\ci{zeus98,zeus96} 
 (open squares) for $\rho$ and $\phi$ production, respectively. 
 The open triangles represent preliminary ZEUS data
 \protect\ci{ZEUS-prel-rho,ZEUS-prel-phi} for $\rho$ and $\phi$
 electroproduction. The solid (dashed) lines 
 are our results for the ratio of differential (integrated) cross
 sections, $\tilde{R}\, (R)$. The ratio $\tilde{R}$ is evaluated at $t=-0.15\,\gev^2$.}
\label{fig:ratio}
\end{center}
\end{figure}

The HERA experiments also measured the decay angular distributions of
the $\rho$ and $\phi$ mesons and determined their spin density matrix
elements. This information allows for a determination of the cross section ratio 
\be
  R(V) \= \frac{\sigma_L(\gamma^*p\to Vp)}{\sigma_T(\gamma^*p\to Vp)}\,,
\label{ratio}
\ee
from which, in combination with \req{int-cross}, the longitudinal
cross section, $\sigma_L$, can be isolated as well. The HERA data for
$\sigma_L$  and $R$ are compared to our results in Figs.\ \ref{fig:sigrhoL} 
and \ref{fig:ratio}, respectively. Again reasonable agreement is
to be observed for $Q^2$ larger than $4\,\gev^2$. The ratio $R$
increases with $Q^2$ since the transverse cross section is
suppressed by $1/Q^2$ as compared to the longitudinal one, see the
hierarchy \req{hierarchy}. 
%\clearpage

The experimental results on cross section ratio are derived from data on the spin
density matrix element $r_{00}^{04}$. The extracted ratio is therefore
a ratio of the differential cross sections \req{cross-lt}
\be 
\tilde{R}(V) \= \frac{d\sigma_L(\gamma^* p\to Vp)}{d\sigma_T(\gamma^*
  p\to Vp)}\,,
\label{R-tilde}
\ee
which equals the ratio of integrated cross sections, $R$, only if both
the differential cross sections show the same $t$ dependence
\footnote{For single exponentials the relation between $R$ and
    $\tilde{R}$ is given by\\ $ R={B_{TT}}/{B_{LL}} 
    \exp{[-(B_{LL}-B_{TT})t]}\, \tilde{R}(t)$.}.
This is however not the case if the slopes differ. Therefore, $\tilde{R}$,
measured at $t\simeq -0.15\,\gev^2$, is about $10-20\%$ larger than
$R$. In Fig.\ \ref{fig:ratio} we also display our prediction for
$\tilde{R}$. Very good agreement with experiment is to be seen now. 
It is to be stressed that the uncertainties of the gluon distribution
\ci{CTEQ} entail a typical error of about $30\%$ for our predictions
for the cross sections. In the ratios these errors cancel to some extent. 
As we remarked a fit with $B^V_{TT}\simeq B^V_{LL}$ is also in
agreement with the present data provided the value of the product
$(f_{V\,T}/M_V)^2/B^V_{TT}$ is kept constant. The ratio $R$ for this
fit practically falls together with $\tilde{R}$ in the fit presented above. 
 
\begin{figure}[t]
\begin{center}
\includegraphics[width=0.44\textwidth,height=0.42\textwidth, 
bb= 31 313 478 747,clip=true]{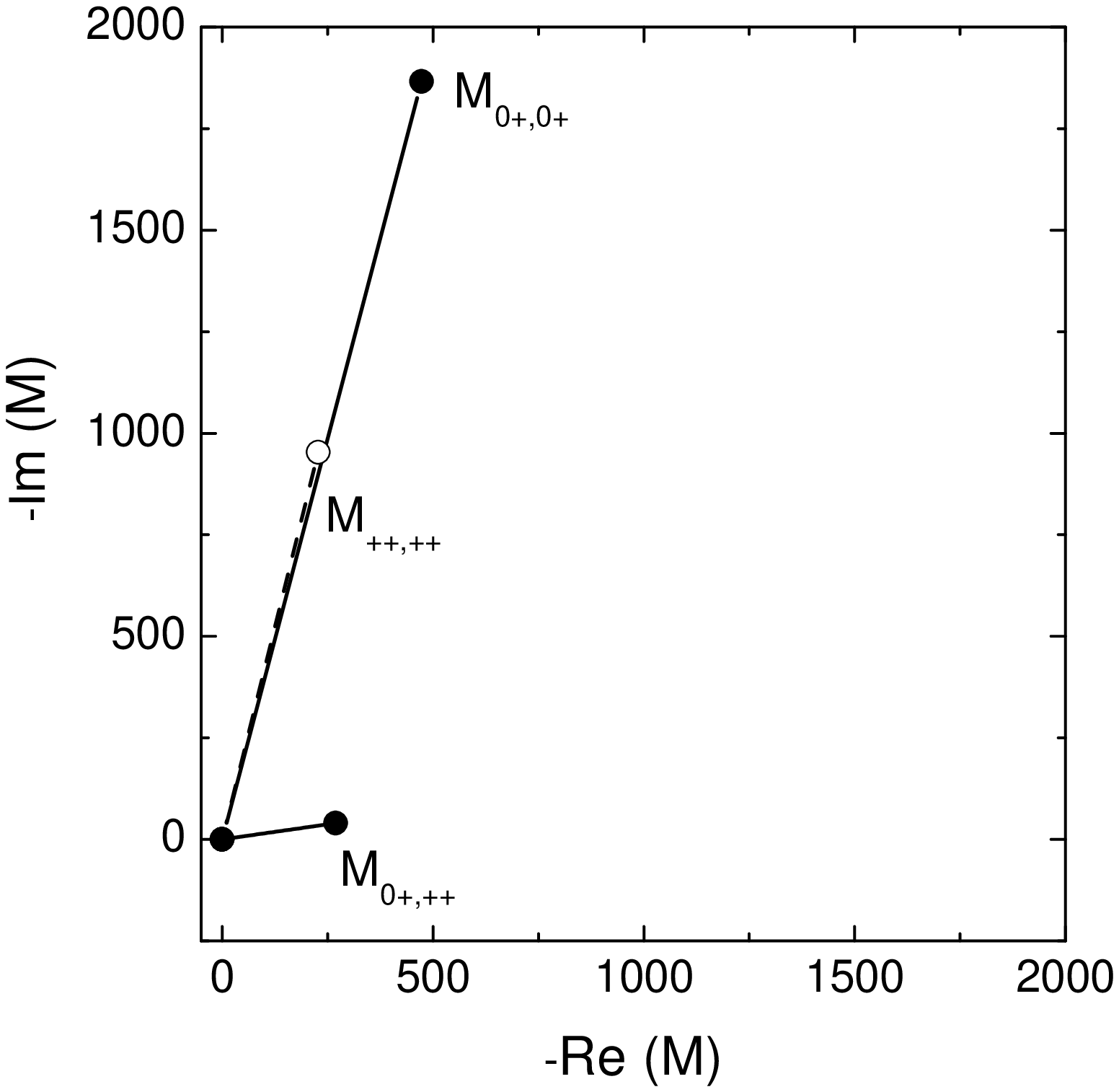}\hspace*{0.3cm}
\includegraphics[width=.42\textwidth,height=0.41\textwidth,
bb=69 317 520 742,clip=true]{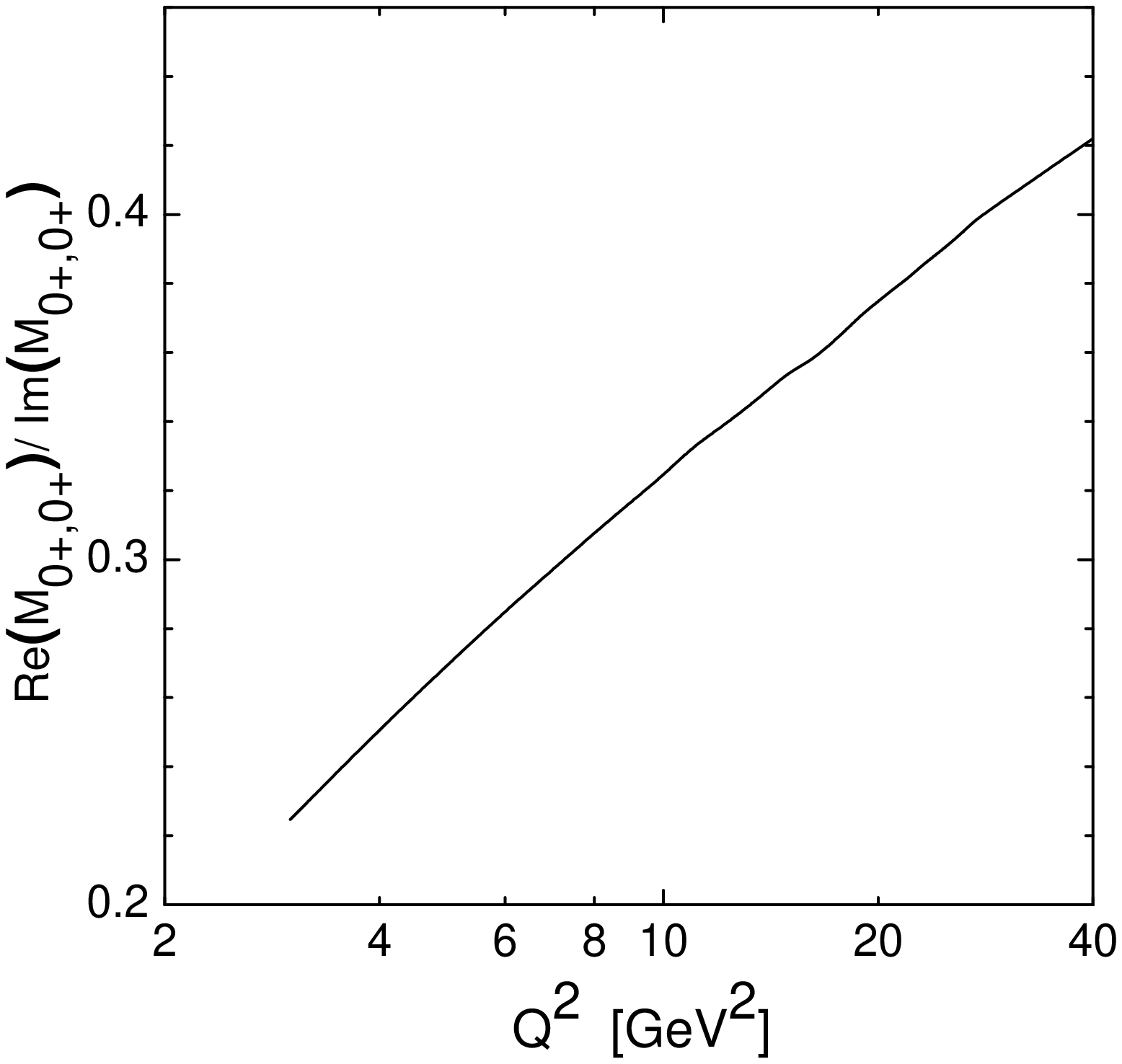}
\end{center}
\caption{Left: The $\rho$ production amplitudes
  for the three transitions $L\to L$, $T\to T$ and $T\to L$ at 
  $Q^2=4\,\gev^2$, $t = -0.15\,\gev^2$ and $W = 75\,\gev$. 
  Right: Real over imaginary part of the amplitude ${\cal
  M}^H_{0+,0+}$ for $\rho$ production versus $Q^2$ at $t\simeq 0$ and 
  $W = 75 \, \gev$.} 
\label{fig:phase}
\end{figure}

In Fig.\ \ref{fig:phase} we display an Argand diagram of the three
forward amplitudes for $\rho$ electroproduction at $Q^2=4 \, \gev^2$, 
$t=-0.15\,\gev^2$ and $W = 75\,\gev$. Both, ${\cal M}^H_{0+,0+}$ 
and ${\cal M}^H_{++,++}$ are dominantly imaginary while the 
$T\to L$ one is nearly real. The latter phase is a consequence of the
branch point of $\sqrt{\xb^2-\xi^2}$ in Eq.\ \req{hard-amp-ji}.
The hierarchy \req{hierarchy} is here seen again.
The phase of the $\rho$-production amplitude 
${\cal M}^H_{0+,0+}$ at $t\simeq 0$ is shown in more detail on the
right hand side of Fig.\ \ref{fig:phase}. The real over imaginary part
ratio increases with $Q^2$ and takes values between 0.2 and 0.4 in the
$Q^2$ region of interest. The real part of the $L\to L$ amplitude
therefore contributes only about $10\%$  to the cross sections. 

A number of comments concerning the leading-twist contribution
\ci{rad96,col96} are in order. As we mentioned above it is given by the
collinear approximation of the dominant amplitude ${\cal M}^H_{0+,0+}$.
The salient features of the
leading-twist contribution are passed on to the full $L\to L$
amplitude, the quark transverse momenta and Sudakov suppressions
essentially affect its absolute value. The examination of the
leading-twist contribution therefore elucidates many properties of our
results in a simple way. Neglecting the $k_\perp$ terms
in \req{propagator} and using the standard definition of a meson \da{} 
\be
 \frac{f_{VL}}{2\sqrt{2N_c}}\, \Phi_{VL}(\tau) \= \int \frac{d^3 \vk}{16\pi^3}\, 
                                                \Psi_{VL}(\tau, k^2_\perp)\,,
\ee
we obtain the subprocess amplitude ${\cal H}^V_{0+,0+}$ in
collinear approximation from Eq.\ \req{tt-ji} and, inserting it into
\req{amp-nf}, the leading-twist contribution to the $L\to L$ amplitude 
\be
{\cal M}_{0+,0+}^{\rm coll} \= e\, \frac{8\pi \als f_{VL}}{N_c Q}\;   
                   \langle 1/\tau \rangle_{VL}\, {\cal C}_V\, 
              \int_0^1 d\xb \,\frac{H^g(\xb,\xi)}{(\xb+\xi)
                    (\xb-\xi + i\hat{\veps})}\,.
\label{eq:glue}
\ee
The $1/\tau$ moment of the meson's \da{} $\Phi_{VL}$ occurring now, 
is denoted by $\langle 1/\tau \rangle_{VL}$. For the \wf{}
\req{wave-l} the associated \da{} is the asymptotic form \req{da-as}
which leads to a value of 3 for the $1/\tau$ moment. 

We can now easily understand the growth of the real over imaginary part
ratio with $Q^2$. Applying the derivative analyticity relation
\ci{bronzan}, frequently but unjustifiedly termed the local dispersion 
relation \ci{eichmann}, to the imaginary part of the leading-twist
amplitude \req{eq:glue}
\be 
{\cal M}^{\rm coll}_{0+,0+} \simeq \Big[i - \frac{\pi}{2} 
          \frac{\partial}{\partial \ln{\xbj}} \xbj \Big]\; 
            {\rm Im}\; {\cal M}^{\rm coll}_{0+,0+}\,,
\ee
and using the low-$\xi$ behavior of the model GPD 
$H^g(\xi,\xi)= \bar{c}_0 (2\xi)^{-\delta}$ (see Eq.\
\req{H-model1}), we find
\be
        {\rm Re}M^{\rm coll}_{0+,0+}/\,{\rm Im} M^{\rm coll}_{0+,0+}\simeq \frac12 \pi\,
        \delta(Q^2)\,.
\ee
The increase of $\delta$ with $Q^2$ ( see Eq.\ \req{kappa}) which has
been calculated by the CTEQ group \ci{CTEQ} with the help of QCD evolution,
is what we see at the right hand side of Fig.\ \ref{fig:phase}.
  
Up to corrections from the real part the integrated longitudinal cross
section reads
\be
\sigma_L^{\rm coll} \= \frac{16\,\pi^{\,4}}{N_c^{\,2}}\,
                       \frac{\ale}{B^{\,V}_{LL}\, Q^{\,6}}\; 
                       \big[\als\, f_{VL} {\cal C}_V\, 
                                        \langle 1/\tau\rangle_{VL} \big]^{\,2}\,
         \left|H^g(\xi,\xi)\right|^{\,2}\,,
\label{sig-coll}
\ee
in collinear approximation. The ratio of the $\phi$ and $\rho$ cross
sections is given by $(f_{\phi L} {\cal C}_\phi/f_{\rho L} {\cal
  C}_\rho)^2$. Our results shown in Figs.\ \ref{fig:sigrho} and
\ref{fig:sigrhoL}, approach this value with increasing $Q^2$. 
Due to the behavior of $H^g(\xi,\xi)$ at small
$\xi$ the cross section behaves as 
\be
\sigma_L^{\rm coll} \propto W^{\;4\delta(Q^2)}\,,
\label{sigmaL}
\ee 
at fixed $Q^2$ and small $\xbj$. The power behavior \req{sigmaL}
comes about as a consequence of the behavior of the GPD and the
underlying gluon distribution. 
We note in passing that in the Regge picture~\ci{regge} the exponent 
$\delta(Q^2)$ is associated with Pomeron exchange.
The intercept of the Pomeron trajectory is related to
$\delta$ by $\alpha_P(0) \= 1 +\delta(Q^2)$. In the Regge model
$\delta$ is a free parameter.

\begin{figure}[t]
\begin{center}
\includegraphics[width=.46\textwidth,bb=28 277 504 749,clip=true]
{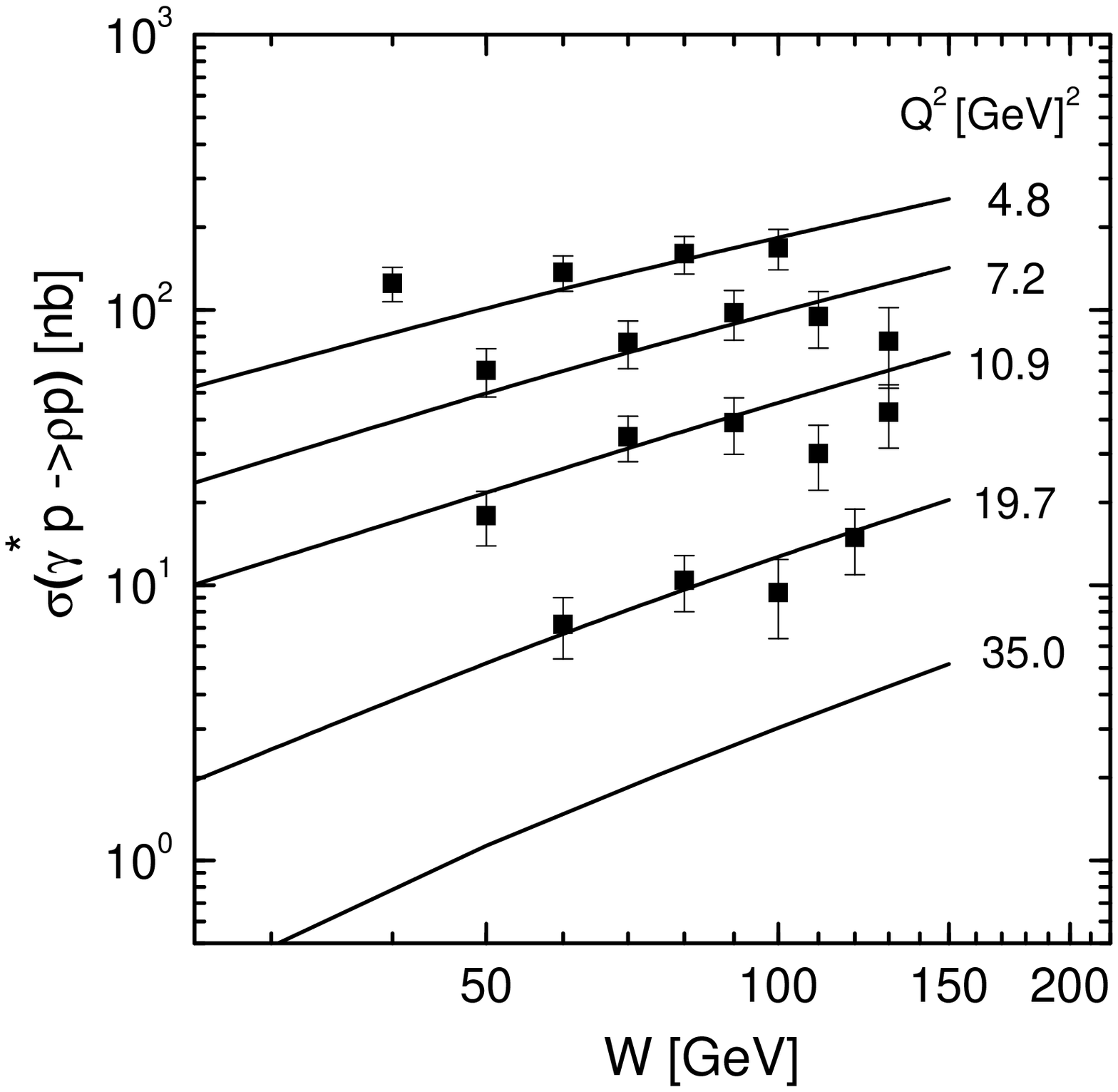} \hspace*{3pt}
\includegraphics[width=0.44\textwidth,bb= 110 310 440 658,clip=true]
{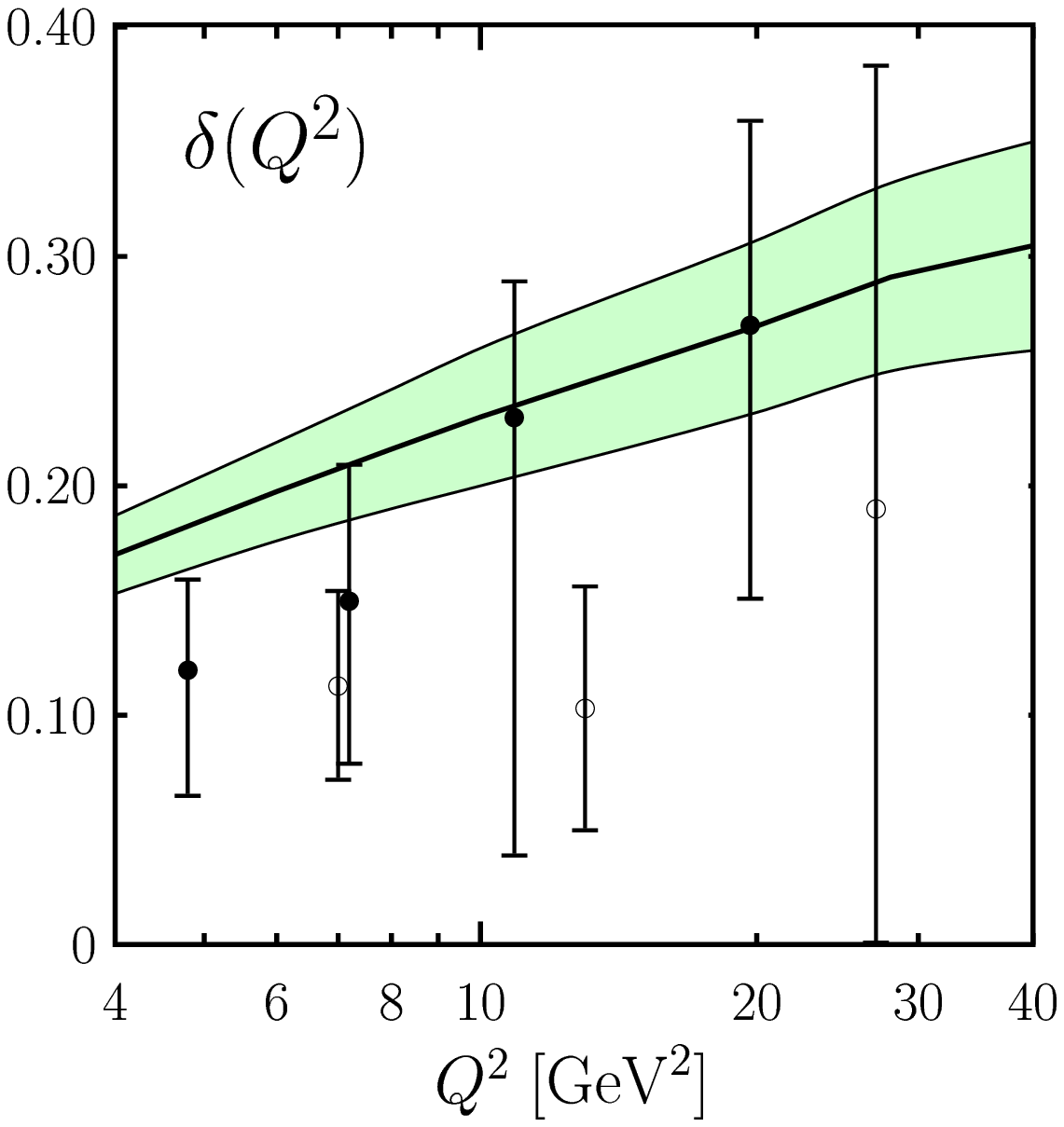}
\end{center}
\caption{Left:The integrated cross section for $\gamma^* p\to \rho p$
 versus $W$ for five values of $Q^2$. Data taken from \protect\ci{h1}. 
 The solid lines represent our result. Right: The power $\delta$
 versus $Q^2$ for $\rho$ electroproduction. The solid curve represents 
 the power as determined in Ref.\ \protect\ci{CTEQ} with an error 
 estimate given by the shaded band. Data are
 taken from \protect\ci{h1} ($\bullet$) and \protect\ci{zeus98} ($\circ$).}
\label{fig:sigrho-W}
\end{figure}

In Fig.\ \ref{fig:sigrho-W} we display the cross section for 
$\gamma^*p\to \rho p$ as a function of $W$ for sample values of
$Q^2$. Fair agreement between experiment and our 
predictions is to be seen. The $W$ dependence of the predictions from
the full approach is very close to that given in \req{sigmaL}.
Deviations from the power law at lower values of $W$, to be observed
in Fig.\ \ref{fig:sigrho-W}, arise from various corrections to the 
leading-twist contribution we take into account, such as the quark 
transverse momenta, the $T\to T$ amplitude and the real parts of the 
$L\to L$ amplitude. This interpretation of the power behavior of
$\sigma_L$ is supported by a comparison of $\delta$ as taken from the
analysis presented in Ref.\ \ci{CTEQ}, with the powers obtained from
fits to the cross section data \ci{h1,zeus98}. Rough
agreement between both the results is to be seen in Fig.\
\ref{fig:sigrho-W} although the errors of the HERA data do not permit
a definite conclusion as yet. Preliminary HERA data seem to improve
the agreement.

For very small $\xi$ one can estimate the size of the collinear
contribution using the leading terms in the model GPD
\req{H-model1}. One obtains
\be
\sigma_L^{\rm coll}(\gamma^*p\to \rho p)= 5.72\, {\rm \mu b} \gev^6\, 
           \left(\frac{\als}{0.3}\right)^2\, 
             \left(\frac{7.5 \gev^{-2}}{B^{\,\rho}_{LL}}\right)\,
             \left(\frac{\bar{c}_0(Q^2)}{2.33}\right)^2\,
            \left(\frac{\langle 1/\tau\rangle_{\rho L}}{3}\right)^2\,  
            \frac{(2\xi)^{-2\delta(Q^2)}}{Q^{\,6}}\,,
\label{cross-num}
\ee
where $\bar{c}_0=c_0/[(1-\delta/3)(1-\delta/2)]$ is the coefficient
of the first term in the power series of $H^g$ \req{H-sum} associated
with Eq.\ \req{dg}. This cross section is rather large, well above
experiment. The quark transverse momenta and the Sudakov factor
suppress it such that agreement with experiment is found, see 
Fig.\ \ref{fig:sigrhoL}.    

\begin{figure}[t]
\begin{center}
\includegraphics[width=.44\textwidth,bb=112 378 437 711,clip=true]{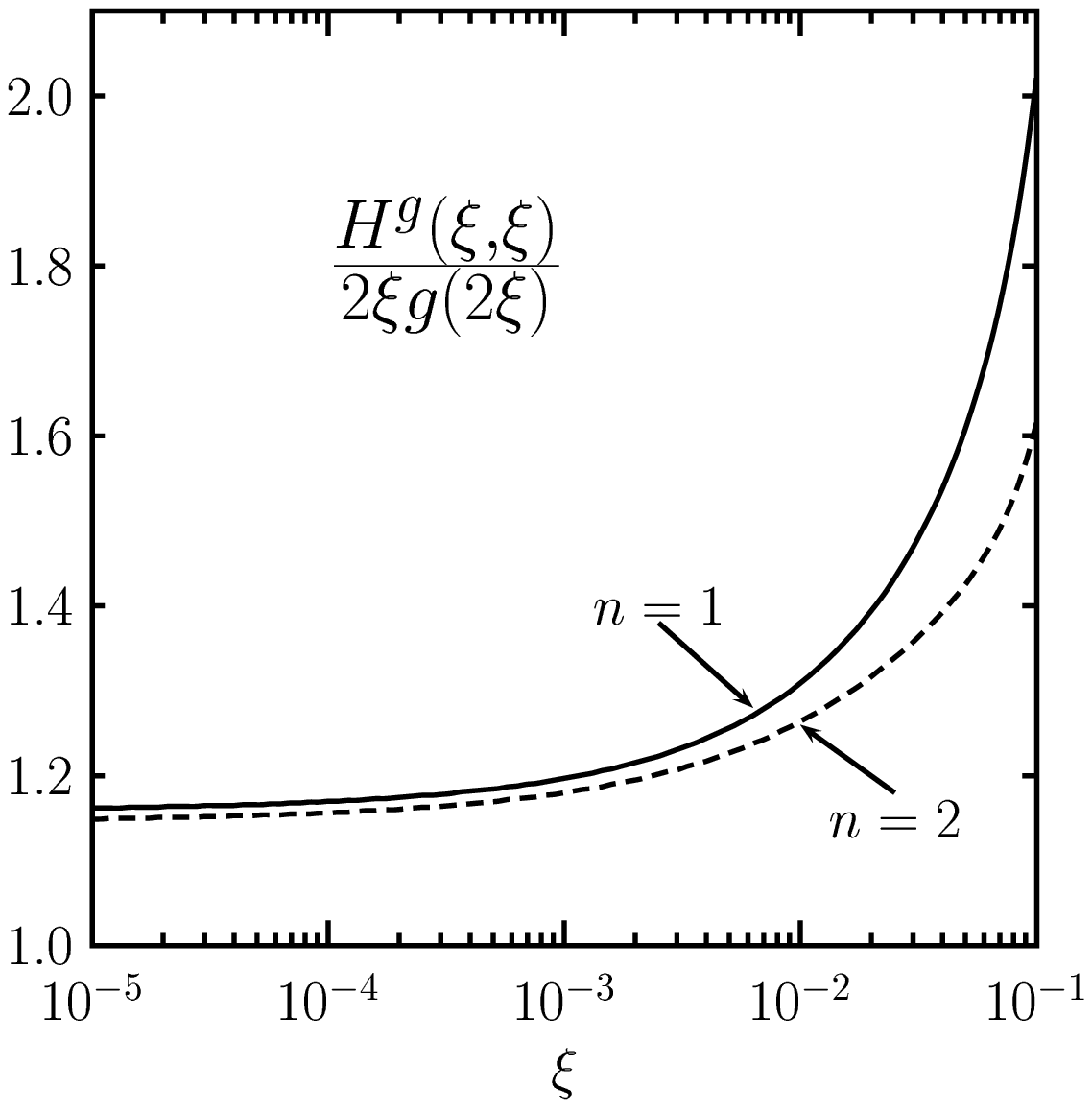}
\end{center}
\caption{Ratios of the GPDs and the parton distribution for the $n=1$ and 2
  models at a scale of $4\,\gev^2$.}
\label{fig:Hg}
\end{figure}

Exploiting the leading $\ln(1/\xbj)$ approximation, Brodsky {\it et al.}
\ci{bro94} found a result~\footnote{Note, in \cite{bro94} the decay 
constant includes the flavor weight factor ${\cal C}_V$.} 
that equals \req{sig-coll} except that $H^g(\xi,\xi)$ is replaced by 
the usual gluon distribution $\xbj g(\xbj)$ (see also \ci{bartels}). 
At very small $\xi$, i.e.\ if $\xi$ is so small that the first terms
in Eq.\ \req{dg} and in the corresponding GPD \req{H-sum} suffice; 
the usual gluon distribution and the GPD only differ by about $20\%$ 
resulting from the difference between $c_0$ and $\bar{c}_0$. For 
larger $\xi$, however, the difference between both the functions
becomes substantial, growing up to about a factor of
1.6 - 2 at $\xi=0.1$, see Fig.\ \ref{fig:Hg}. The use of the 
$\ln(1/\xbj)$ approximation at values of $\xi$ around 0.1 may
therefore lead to an underestimate of the gluonic contribution to
cross sections by a factor 3 to 4. We repeat that, in contrast to 
the $\ln(1/\xbj)$ approximation, we do not require small $\xi$ in 
principal. We only restrict ourselves to small $\xi$ in order to avoid
complications with potential contributions from quarks emitted and
reabsorbed by the proton. The enhancement effect
apparent in Fig.\ \ref{fig:Hg}, is known as the skewing effect
and has been discussed by several authors \ci{mrt,man98,martin97,bel02}. 
The size of the skewing effect estimated in these papers, is 
compatible with our model result for $\xi\ll 1$.        

In any case the leading-twist as well as the $\ln(1/\xbj)$ result provide
cross sections that are too large. In order to settle this problem for the
$\ln(1/\xbj)$ approximation Frankfurt {\it et al} \ci{fra95} estimated 
a correction factor by allowing for quark transverse momenta in the
loop. This mechanism bears resemblance to our approach as we mentioned
in Sect.\ \ref{sec:sub}. The correction factor obtained in \ci{fra95} 
is large enough to achieve agreement with experiment. This factor has 
also been used by Mankiewicz {\it et al.} \ci{man98} in an explorative 
study of $\sigma_L$ in an otherwise collinear GPD approach. Martin 
{\it et al.} \ci{mrt} exploited the $\ln(1/\xbj)$ approximation in
their analysis of vector-meson electroproduction by including
parton transverse momenta and an unintegrated gluon distribution.

%%%%%%%%%%%%%%%%%%%%%%%%%%%%%%%%%%%%%%%%%%%%%%%%%%%%%%%%%%%%%%%%
\section{Spin density matrix elements}
\label{sec:sdme}
%%%%%%%%%%%%%%%%%%%%%%%%%%%%%%%%%%%%%%%%%%%%%%%%%%%%%%%%%%%%%%%%
With the help of Eqs.\ \req{parity}, \req{hierarchy} and \req{npe} 
the spin density matrix elements extracted from the decay angular
distributions measured with unpolarized leptons and protons \cite{schi},
simplify to ($\tilde{R}$ is defined in \req{R-tilde}) 
\ba
N_L&=& 2\, \big|\;{\cal M}^H_{0\,+,0\,+}\big|^2\,,\nn\\[0.3em]
N_T&=& 2 \left[\, \big|{\cal M}^H_{++,++}\big|^2 + 
     \big|{\cal M}^H_{0\,+,++}\big|^2\,\right]\,,\nn\\[0.3em] 
r_{00}^{04} &=& \frac1{1 + \veps \tilde{R}}
           \Big[\,\frac{2}{N_T}\, \big|{\cal M}^H_{0+,++}\big|^2  
                                     + \veps \tilde{R}\,\Big]\,, \nn\\[0.3em] 
{\rm Re}\;r_{10}^{04}&=& - {\rm Re}\; r^1_{10} \,=\, {\rm Im}\; r^2_{10}\,
                     =\, \frac1{1 + \eps \tilde{R}}\, \frac1{N_T}\, {\rm Re}\; 
                     \Big[\,{\cal M}^H_{++,++}\; 
                     {\cal M}^{H*}_{0\,+,++}\,\Big]\,,\nn\\[0.3em]
r_{00}^1&=&\frac{-1}{1 + \eps \tilde{R}}\, \frac{2}{N_T} 
                    \big|{\cal M}^H_{0\,+,++}\big|^2\,, \nn\\[0.3em]
r_{1-1}^1 &=& -{\rm Im}\; r_{1-1}^2
  =\frac1{1 + \eps \tilde{R}}\,\frac{1}{N_T}\, \big|{\cal M}^H_{++,++}\big|^2
                                                            \,,\nn\\[0.3em]
r_{00}^5&=& \frac{4}{\sqrt{2N_L\,N_T}}\,\frac{\sqrt{\tilde{R}}}{1 + \eps \tilde{R}}\, 
{\rm Re}\; \Big[\,{\cal M}^H_{0\,+,0\,+}\; 
                     {\cal M}^{H*}_{0\,+,++}\,\Big] \,,\nn\\[0.3em]   
{\rm Re}\;r_{10}^5 &=& - {\rm Im}\;r_{10}^6 =  
            \frac{\sqrt{\tilde{R}}}{1 + \eps
               \tilde{R}}\,\frac{1}{\sqrt{2N_L\,N_T}}\,  
       {\rm Re}\; \Big[\, {\cal M}^H_{++,++}\; 
                   {\cal M}^{H*}_{0\,+,0\,+}\,\Big] \,,\nn\\[0.3em]
\label{density}
\ea
while 
\be
r_{1-1}^{04} = r_{11}^1=r_{11}^5= r_{1-1}^5 =
                                      {\rm Im}\; r_{1-1}^6 =0\,,
\label{equal}
\ee
because of the neglect of $L\to T$ and $T\to -T$ transitions. 
The relations \req{density}, obtained in the GPD approach under the
assumption of the dominance of the $H^g$ terms, coincide with those
found assuming dominance of natural parity $t$-channel exchanges 
and the neglect of proton helicity flip~\ci{h1,schi}. 
The contributions from $\widetilde{H}^g$ enter
the spin density matrix elements only as bilinears, there are no
interferences between $H^g$ and $\widetilde{H}^g$ terms.

\begin{figure}[p,h]
\begin{center}
%\includegraphics[width=0.78\textwidth, bbllx=80pt,bblly=493pt,
%bburx=520pt,bbury=715pt,clip=true]{sdmerho1.ps}
%\includegraphics[width=0.78\textwidth, bbllx=85pt,bblly=438pt,
%bburx=524pt,bbury=705pt,clip=true]{sdmerho2.ps}
\includegraphics[width=0.78\textwidth, bbllx=0pt,bblly=0pt,
bburx=436pt,bbury=488pt,clip=true]{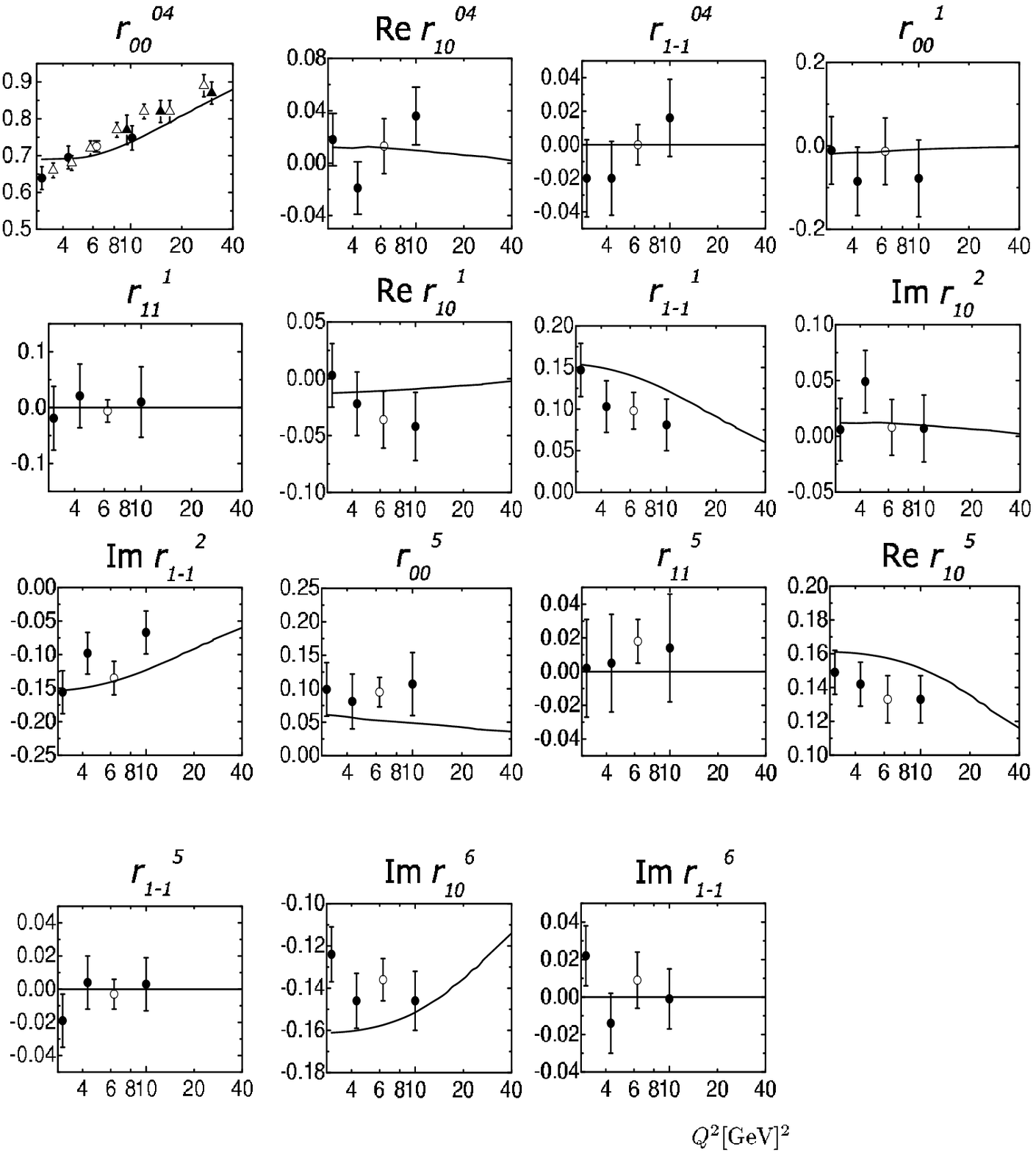}
\end{center}
\caption{The spin density matrix elements of electroproduced  $\rho$
  mesons versus $Q^2$ at $W\simeq 75\, \gev$ and $t \simeq -0.15
  \,\gev^2$. Data, taken from \protect\ci{h1} (filled circles) and
  \protect\ci{zeus99} (open circles), are compared to our results
  (solid line). Preliminary data on $r_{00}^{04}$ from
  ZEUS \protect\ci{ZEUS-prel-rho} (open triangles) are also shown.}
\label{fig:Srho}
\end{figure}

\begin{figure}[p]
\begin{center}
\includegraphics[width=0.78\textwidth,bbllx=80pt,bblly=227pt,
bburx=518pt,bbury=714pt,clip=true]{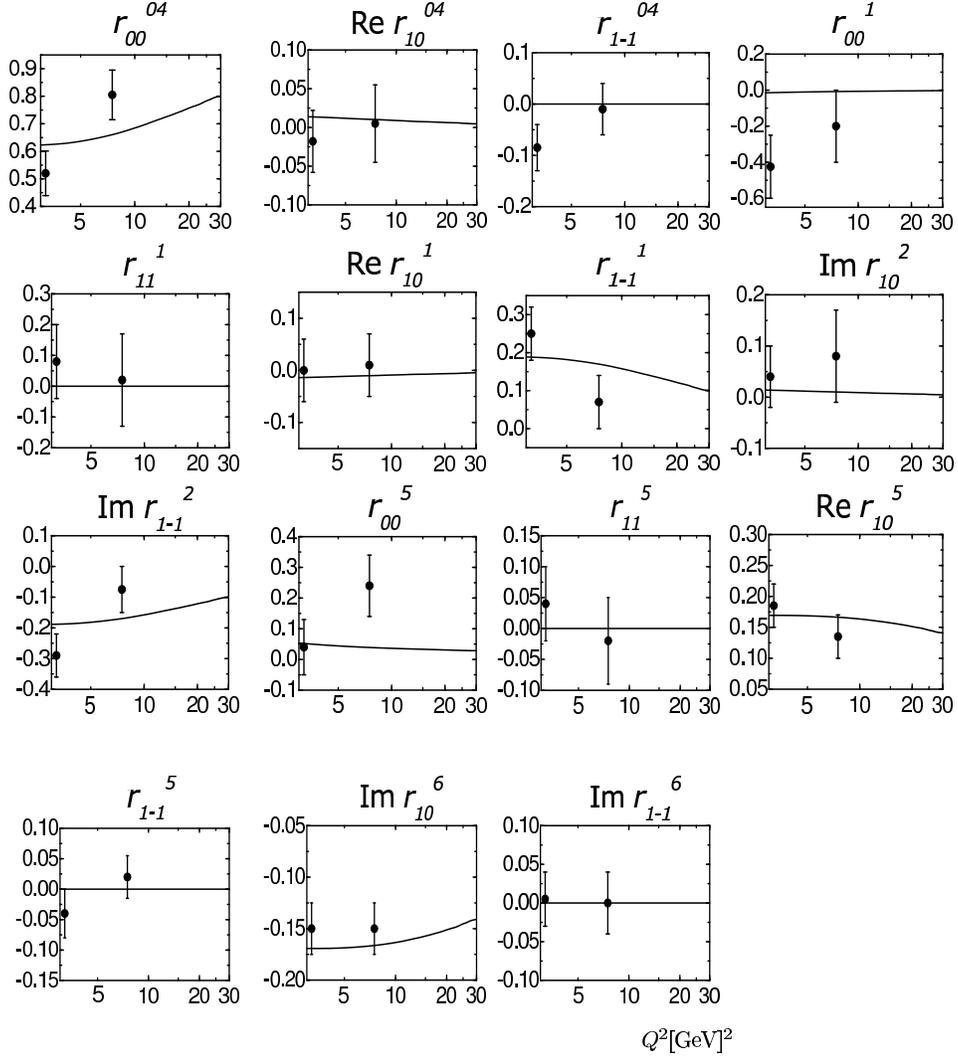}
\end{center}
\caption{The spin density matrix elements for $\phi$ electroproduction 
versus $Q^2$ at $W \simeq 75\,\gev$ and $t \simeq -0.15\,\gev^2$. The
 H1 data \protect\ci{adloff} are compared to our results (solid line).} 
\label{fig:sdmephi}
\end{figure}

The data for the spin density matrix elements from H1 \ci{h1,adloff} 
and ZEUS \ci{zeus99}, are shown in Figs.\ \ref{fig:Srho} and \ref{fig:sdmephi}
and compared to the results from the GPD based approach. The general
pattern of the data is reproduced. The dominance of the $L\to
L$ transition amplitude is clearly visible in the angular distribution
of the production and decay of the vector mesons, in particular in the
value of $r_{00}^{04}$ which tends towards 1 with increasing $Q^2$. 
This behavior is well reproduced by our approach as we already 
discussed in connection with the cross section ratio $R$. 

The $T\to L$ amplitude is probed by the matrix elements $r_{00}^1$ and
$r_{00}^5$. While the first matrix element is approximately  
$\propto |{\cal M}^H_{0+,++}|^2/|{\cal M}^H_{0+,0+}|^2$, the ratio
${\rm Im}\,{\cal M}^H_{0+,++}/{\rm Im}\,{\cal M}^H_{0+,0+}$ 
essentially controls the second. Both $r_{00}^5$ and 
$|r_{00}^1|$ are found to be rather small. The ratio of the $T\to T$ 
and $L\to L$ amplitudes is approximately measured by 
$r_{1-1}^1$ and ${\rm Re}\; r_{10}^5$, quadratically in the first case,
linearly is the second one since both the amplitudes have about
the same phase as is shown in Fig.\ \ref{fig:phase}. The fair
agreement between theory and experiment for these spin density matrix
elements tells us that our approach provides the correct sizes and
relative phases of the $T\to T$ and $L\to L$ amplitudes.  
The matrix elements ${\rm Re}\; r_{10}^{04}=-{\rm Re}\; r_{10}^1= {\rm Im}\;
r_{10}^2$ measure an interference term between the $T\to T$ and $T\to
L$ amplitudes which is very small. Also this prediction is in 
acceptable agreement with experiment.

\begin{figure}[p,h]
\begin{center}
%\includegraphics[width=0.78\textwidth, bbllx=80pt,bblly=493pt,
%bburx=520pt,bbury=715pt,clip=true]{sdmerhobt1.ps}
%\includegraphics[width=0.78\textwidth, bbllx=85pt,bblly=438pt,
%bburx=524pt,bbury=705pt,clip=true]{sdmerhobt2.ps}
\includegraphics[width=0.78\textwidth, bbllx=80pt,bblly=225pt,
bburx=521pt,bbury=715pt,clip=true]{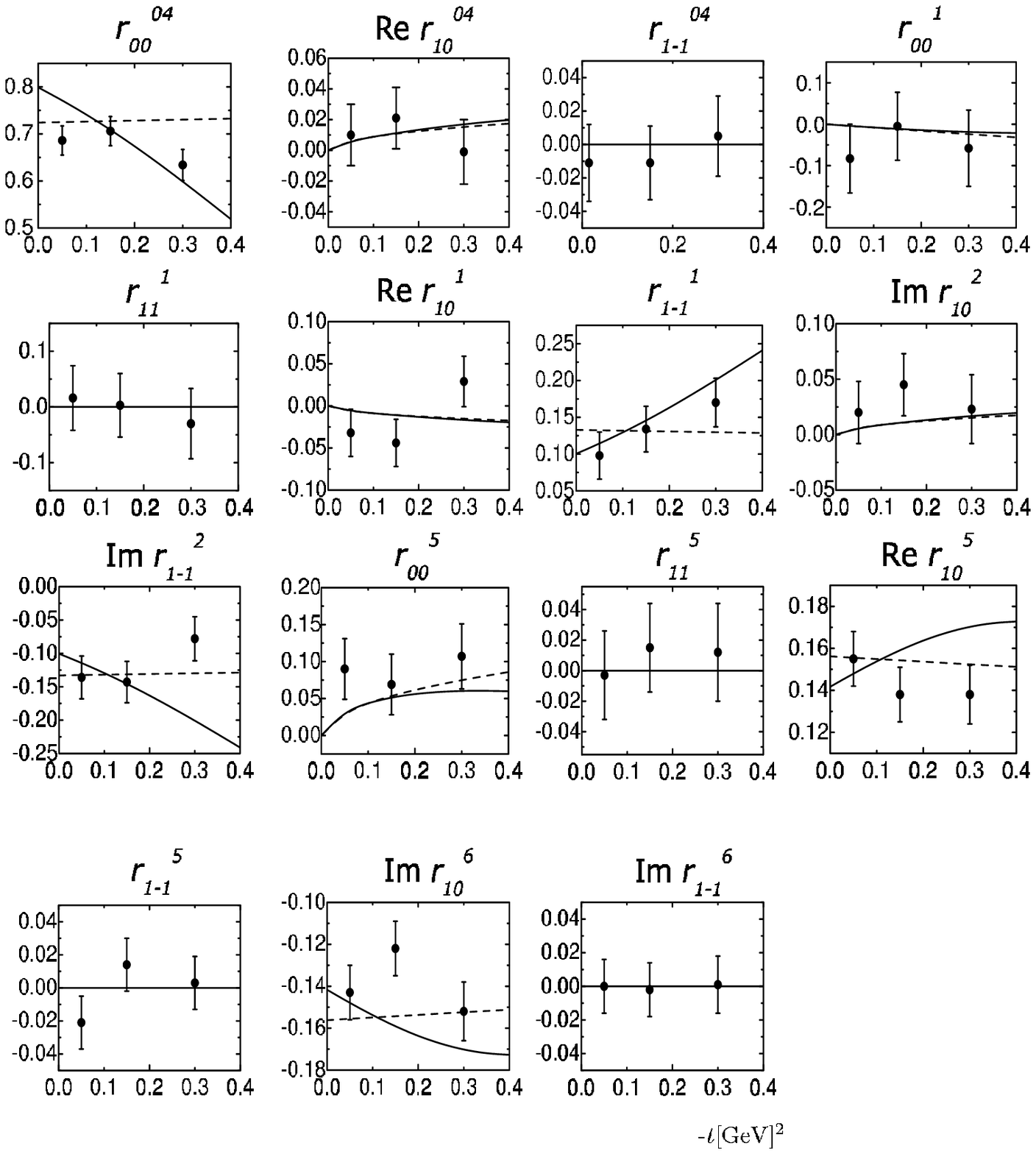}
\end{center}
\caption{The spin density matrix elements of electroproduced  
$\rho$ mesons versus $t$ at $Q^2=5\,\gev^2$ and  $W \simeq 75\,\gev$.
Data taken from \protect\ci{h1} (filled circles). The solid (dashed) lines
represent our results for the choice $B^V_{TT}=B^V_{LL}/2\; (B^V_{LL})$.}
\label{fig:sdme-t}
\end{figure}

The $t$ dependence of the spin density matrix elements confirm the
above observations, see Fig.\ \ref{fig:sdme-t}. The $T\to L$ sensitive 
matrix elements behave proportional to $\sqrt{-t}$ or $t$ while those 
controlled by ratios of the $T\to T$ and $L\to L$ amplitudes exhibit 
an $t$ dependence according to the different slopes chosen for them.  
As we mentioned in Sect.\ \ref{sec:cross} the freedom in choosing a
suitable value of $M_V$ also allows fits with $B^V_{TT}\simeq
B^V_{LL}$. While the transverse cross section is nearly insensitive to
this choice provided the product $(f_{V\,T}/M_V)^2/B^V_{TT}$ is
approximately kept fixed, does the $t$ dependence of some of the spin
density matrix elements (e.g.\ $r_{00}^{04}$, $r_{1-1}^{1}$) change;
they become very flat in $t$. Given the accuracy of the present data
\ci{h1} such a behaviour is not in conflict with experiment.
% although
%the fit with $B^V_{TT}=B^V_{LL}/2$ leads to a somewhat better
%agreement with the data.

\begin{figure}[p,h]
\begin{center}
\includegraphics[width=0.78\textwidth, bbllx=80pt,bblly=227pt,
bburx=517pt,bbury=716pt,clip=true]{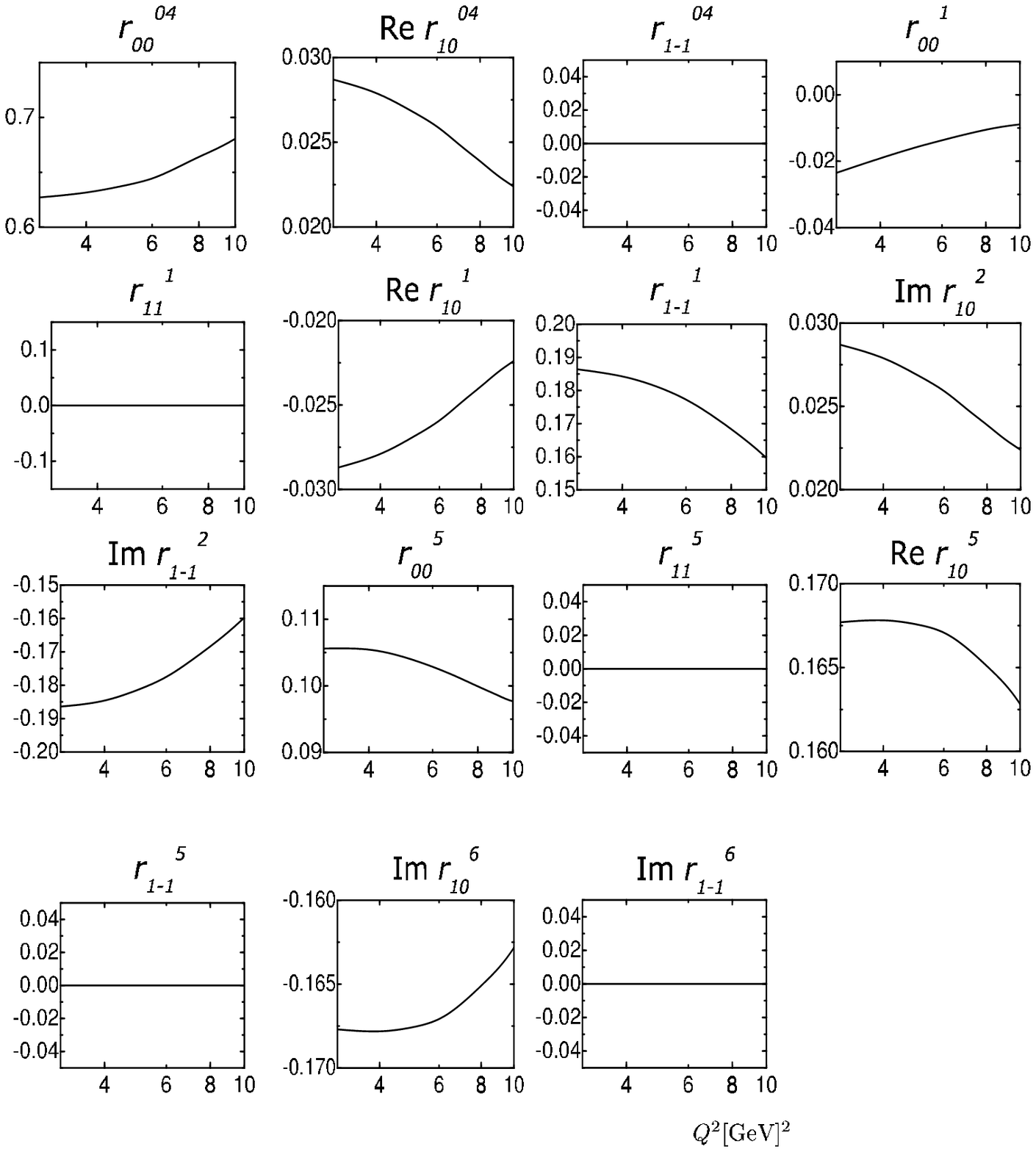}
\end{center}
\caption{The spin density matrix elements of electroproduced $\phi$ 
mesons versus $Q^2$ at $W\simeq 10\,\gev$, $y\simeq 0.6$ and 
$t \simeq -0.15\,\gev^2$. The solid lines represent our results.}
\label{fig:sdme-low}
\end{figure}

Finally, in Fig.\ \ref{fig:sdme-low} we show our predictions for
$\phi$ electroproduction at $W=10\,\gev$ characteristic of the
COMPASS experiment. 

Other theoretical analyses \ci{iva98,nem94,roy00,iva03} of the spin density
matrix elements base on variants of the $\ln(1/\xbj)$ approximation.
The variants differ from each other in the detailed
treatment of the subprocess $\gamma^* g\to q\bar{q} g$. The same
hierarchy of the amplitudes are obtained as we find and, in general, rather
similar results are obtained for the spin density matrix elements. Worth
mentioning is the different phase of the $T\to L$ amplitude and a
somewhat different $t$ dependence of the matrix elements.  

%%%%%%%%%%%%%%%%%%%%%%%%%%%%%%%%%%%%%%%%%%%%%%%%%%%%%%%%%%%%%%%%%
\section{The helicity correlation}
\label{sec:all}
\begin{figure}
\begin{center}
\includegraphics[width=0.48\textwidth, bb= 33 365 538 742,clip=true]{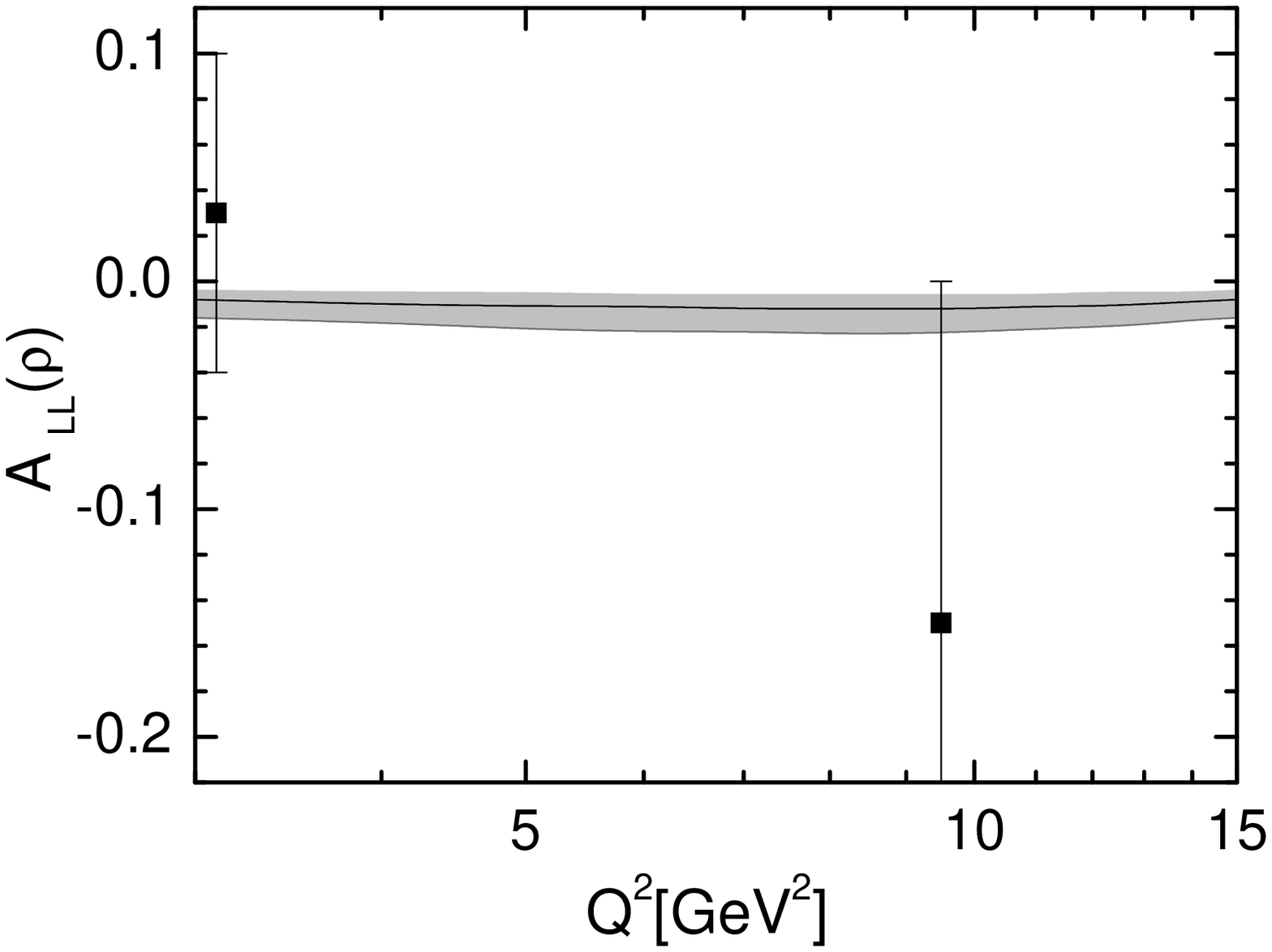}
\includegraphics[width=0.47\textwidth, bb= 16 348 533 742,clip=true]{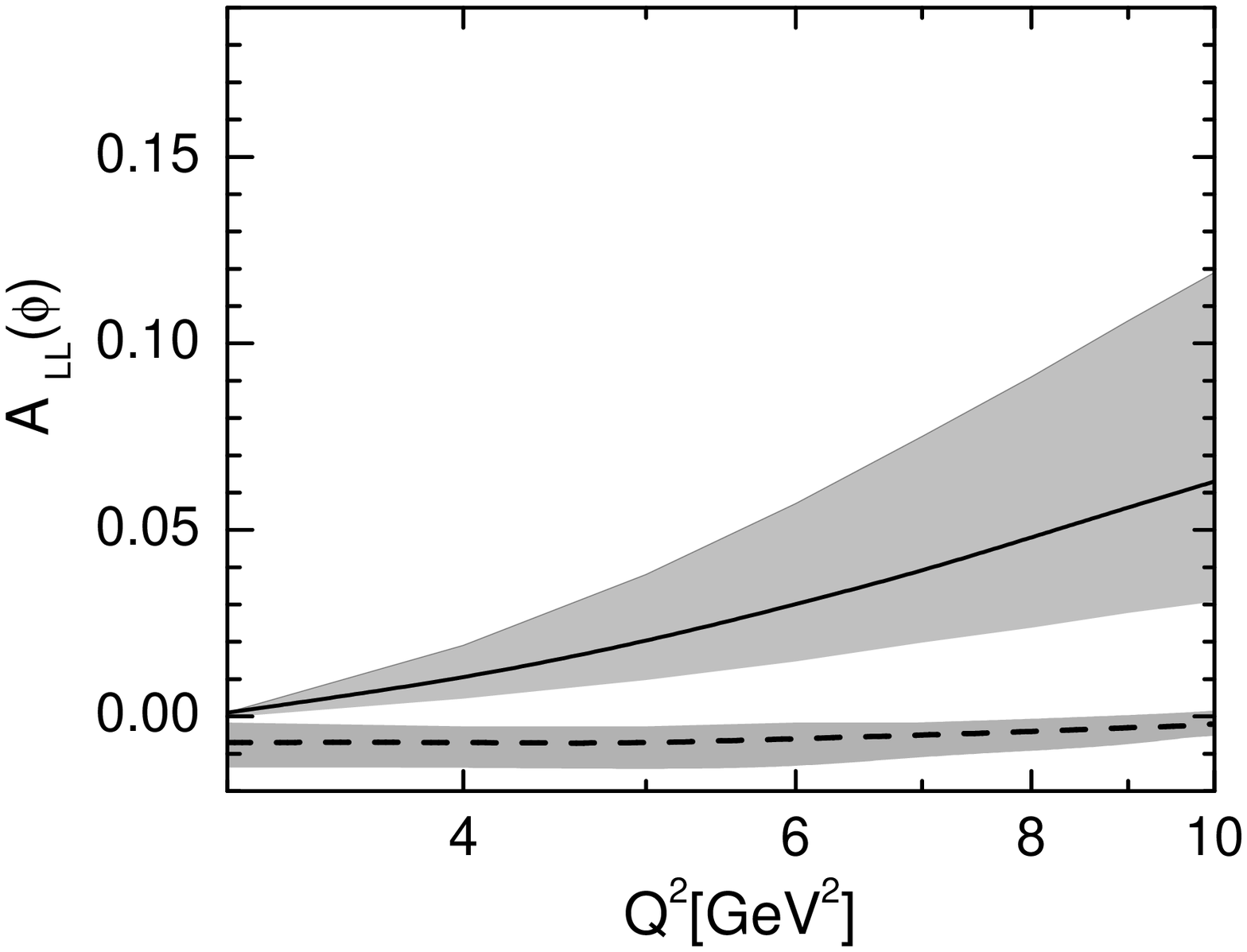}
\end{center}
\caption{Left: The helicity correlation $A_{LL}$ for $\rho$
  electroproduction versus $Q^2$ at $W=15\,\gev$, $t\simeq 0$ and
  $y\simeq 0.6 $. Data taken from SMC \protect\ci{compass}. Right:
  $A_{LL}$ for $\phi$ production at $W=5\,\gev$ (solid line) and 
  $W=10\,\gev$ (dashed line); $y\simeq 0.6$. The shaded bands reflect 
  the uncertainties in our predictions due to the error in the 
  polarized gluon distribution \ci{BB}.}
\label{fig:all}
\end{figure}

%%%%%%%%%%%%%%%%%%%%%%%%%%%%%%%%%%%%%%%%%%%%%%%%%%%%%%%%%%%%%%%%%%%
Last we want discuss the role of the GPD $\widetilde{H}^g$. For this
purpose we consider the initial state helicity correlation $A_{LL}$
which can be measured with longitudinally polarized beam and target. 
After integration over the azimuthal angle this correlation reads
\be
A_{LL}[ep\to epV]\= \frac{\sqrt{1-\veps^2}}{32\pi W^2(W^2+Q^2)}\,
                       \frac{\big|{\cal M}_{++,++}\big|^2 
              + \big|{\cal M}_{0\,+,++}\big|^2 - \big|{\cal M}_{-+,-+}\big|^2
               - \big|{\cal M}_{0\,+,-+}\big|^2}
                 {d\sigma_T/dt + \veps d\sigma_L/dt}\,,
\label{all}
\ee
where the amplitudes and cross sections refer to the process
$\gamma^*p\to Vp$ and are given in Eqs.\ \req{amp-nf} and \req{cross-lt}. 
As can easily be seen from Eq.\ \req{npe} $A_{LL}=0$ if the
$\widetilde{H}^g$ terms are neglected as we did in the preceding
sections. Yet in contrast to the cross sections and spin density
matrix elements where the correction are bilinear in the
$\widetilde{H}^g$ terms and, hence, extremely small, the leading term 
in $A_{LL}$ is an interference between
the $H^g$ and the $\widetilde{H}^g$ terms. In fact, with the help of
Eqs.\ \req{parity} and \req{npe}, one obtains from Eq.\ \req{all}
\be
 A_{LL}[ep\to epV]\= 2 \sqrt{1-\veps^2}\,
                \frac{{\rm Re}\;\Big[ {\cal M}^H_{++,++}\, 
                 {\cal M}^{\widetilde{H}*}_{++,++}\Big]}
               {\veps |{\cal M}^H_{0+,0+}|^2 + |{\cal M}^H_{++,++}|^2}\,. 
\ee
Obviously, this ratio is of order $\langle k^2_\perp\rangle/Q^2\;\langle
\widetilde{H}^g\rangle/\langle H^g \rangle$ and, therefore, very small
values for $A_{LL}$ are to be expected. Indeed exploiting the model GPDs
presented in Sect.\ \ref{sec:gpd} we confirm this assertion as can be
seen from Fig.\ \ref{fig:all} where results for $A_{LL}$ for $\rho$
and $\phi$ electroproduction at $t\simeq 0$ are displayed. The results
for $\rho$ production, only shown at $W=15\, \gev$, is compared to the 
SMC data \ci{compass}. At this energy and in the range of $Q^2$ shown
in the plot, the contribution from the quark GPD is expected to be
small \ci{diehl04}. Our results for $A_{LL}$ are not in disagreement 
with experiment given the admittedly large experimental errors and the 
rather large value of the skewness. Results for $\phi$
electroproduction are shown at energies typical for the HERMES and 
COMPASS experiments. The dominance of the gluon over the sea quarks 
permits this. At $W=5\,\gev$ $A_{LL}$ is not very small since the major 
contribution to it comes from the region $0.1 \lsim \xb \lsim 0.2$  
where $\Delta g/g$ is not small. 

The proton helicity flip contribution, related to the GPD $E^g$, may change 
these results but likely not substantially. The helicity correlation
will increase with growing momentum transfer if the slope of the $T\to
T$ amplitude is smaller than that of the $L\to L$ one.
Besides allowing predictions for $A_{LL}$  this calculation
also supports our assumption of negligible contributions from 
$\widetilde{H}^g$ to cross sections and spin density matrix elements.
%%%%%%%%%%%%%%%%%%%%%%%%%%%%%%%%%%%%%%%%%%%%%%%%%%%%%%%%%%%%%%%%
\section{Summary}
%%%%%%%%%%%%%%%%%%%%%%%%%%%%%%%%%%%%%%%%%%%%%%%%%%%%%%%%%%%%%%%%
We analyzed electroproduction of light vector mesons at small $\xbj$ 
within a GPD based approach. In this kinematical domain the gluonic
GPD $H^g$, parameterizing the response of the proton to the emission and
reabsorption of gluons, controls the process. The gluonic GPD, not
calculable at present, is constructed from an ansatz for the double
distributions currently is use. In order to examine the influence of
the model GPD on the numerical results for vector meson
electroproduction we used two different versions for it ($n=1$ and
2). The differences in the numerical results obtained from these two
models, are on the percent level. The subprocess amplitudes for 
$\gamma^* g\to Vg$ are calculated by us to lowest order of
perturbative QCD but transverse momenta of the quark and antiquark
that form the vector meson, are taken into account as well as Sudakov
suppression which sum up gluonic radiative corrections. 

The GPD approach reproduce all main features of vector meson
electroproduction at small $\xbj$ known from phenomenology. The
dominance of the contributions from the GPD $H^g$ over those from
$\widetilde{H}^g$ and $E^g$, leads to the relations \req{npe} and,
hence, to results equivalent to those obtained assuming the dominance
of natural parity exchange. Approximate $s$-channel helicity
conservation holds due to the hierarchy \req{hierarchy} the amplitudes
respect in our GPD based approach. The behavior of the longitudinal
cross section as a $Q^2$ dependent power of $W$ at fixed $Q^2$ is a
consequence of low $\xi$ properties of the GPD and QCD evolution. 
The numerical results we obtain from our approach are in reasonable
agreement with the small $\xbj$ data on cross sections and spin
density matrix elements for electroproduction of $\rho$ and $\phi$
mesons measured by H1 and ZEUS. The $t$ dependence of vector meson
electroproduction is not yet satisfactorily settled. In principle it is
generated by a combination of the $t$ dependence of the GPD and, with
lesser importance, that of the subprocess amplitudes. Due to the lack of a
plausible parameterization of the $t$ dependence of the GPD we have
evaluated the electroproduction amplitudes at $t\simeq 0$ and
multiplied them by exponentials in $t$. Improvements on this recipe
are demanded and will be unavoidable as soon as detailed differential
cross section data are at hand.

We also compared in some detail our approach to the leading-twist
contribution and to the leading $\ln(1/\xbj)$ approximation. The latter is
rather close to the GPD approach at low $\xbj$ and small $t$ but not
identical. For $\xbj$ larger than about 0.01 the replacement of 
$H^g(\xi,\xi)$ by $2\xi g(2\xi)$ becomes inappropriate. The GPD has,
in contrast to the leading  $\ln(1/\xbj)$ approximation, the potential
to investigate the $t$ dependence of electroproduction. The lack of
understanding of the GPD's $t$ dependence prevents this at present.    

%%%%%%%%%%%%%%%%%%%%%%%%%%%%%%%%%%%%%%%%%%%%%%%%%%%%%%%%%%%%%%%%
\section*{Acknowledgements}
%%%%%%%%%%%%%%%%%%%%%%%%%%%%%%%%%%%%%%%%%%%%%%%%%%%%%%%%%%%%%%%%%
We thank A. Borissov,  A. Efremov,  O. Nachtmann, H. Meyer, E. Paul, 
A. Sandacz and L. Szymanowski for discussions and A. Bruni for
providing us with preliminary ZEUS data. This work has been supported 
in part by the Russian Foundation for Basic Research, 
Grant 03-02-16816, the Integrated Infrastructure Initiative 
``Hadron Physics'' of the European Union, contract No. 506078 and 
by the Heisenberg-Landau program.
%%%%%%%%%%%%%%%%%%%%%%%%%%%%%%%%%%%%%%%%%%%%%%%%%%%%%%%%%%%%%%%%

\end{document}